\documentclass[preprint,times,10pt]{elsarticle}
\usepackage[left=1in,right=1in]{geometry}

\usepackage{amssymb}
\usepackage{amsmath}
\numberwithin{equation}{section}
\usepackage{physics}
\usepackage{bm}
\usepackage[colorlinks=true]{hyperref}

\biboptions{sort&compress}

\journal{Annals of Physics}

\begin{document}

\begin{frontmatter}



\title{Path Integral Formalism for Quantum Open Systems}


\author{Ruofan Chen} 

\affiliation{organization={College of Physics and Electronic Engineering, and Center for Computational Sciences, Sichuan Normal University},
            city={Chengdu},
            postcode={610068}, 
            country={China}}

\begin{abstract}
  This article provides a detailed derivation of the path integral
  formalism for both boson and fermion quantum open systems using
  coherent states. The formalism on the imaginary-time axis, Keldysh
  contour, and Kadanoff contour are given. The corresponding generating
  functional technique, which can be used to retrieve the environment
  information from the system correlation function, is also discussed.
\end{abstract}


  
  


\end{frontmatter}




\tableofcontents
\section{Introduction}\label{sec:intro}

Path integral formalism is one of the most powerful tools to study
quantum open systems. The path integral formalism was proposed by
Feynman \cite{feynman1948-space}, by which he showed that the
evolution operator of a quantum system can be represented as a sum
over all possible paths in configuration space. Later he and Vernon
\cite{feynman1963-the} developed the general path integral formalism
of a quantum open system with a boson environment. The derivations of
the formalism can be found in several textbooks
\cite{feynman1965-quantum,feynman1972-statistical,schulman1981-techniques,kleinert2006-path,weiss1993-quantum}. The
path integral formalism serves as a cornerstone for studying the
quantum open system. For instance, the path integral expression was
used to study the quantum dissipative effect
\cite{caldeira1981-influence,caldeira1983-quantum}, quantum Brownian
motion \cite{caldeira1983-path}, the phase transition
\cite{bray1982-influence} and the dynamics of the spin-boson model
\cite{leggett1987-dynamics}. The analytical treatment of the path
integral formalism of quantum open system is comprehensively reviewed
in the textbook by Weiss \cite{weiss1993-quantum}. In the 1990s, the
quasi-adiabatic propagator path integral (QuAPI) method
\cite{makarov1994-path,makri1995-numerical}, a numerical exact
algorithm based on tensor multiplication, is proposed to evaluate the
path integral expression. The QuAPI method has been widely used to
study the dynamics of the spin-boson model
\cite{makarov1995-control,makarov1995-stochastic,thorwart2000-iterative,shao2002-iterative,dong2004-quantum,nalbach2009-landau}. Recently,
the time-evolving matrix product operator (TEMPO) method
\cite{strathearn2018-efficient,strathearn2020-modelling} recasts the
tensor multiplication in QuAPI into the modern tensor network
language, which greatly boosts the computational efficiency. The TEMPO
method is now the state-of-art to study the quantum open system with a
boson environment
\cite{joergensen2019-exploiting,fux2021-efficient,popovic2021-quantum,gribben2021-using,gribben2022-exact,otterpohl2022-hidden,chen2023-non,chen2023-heat,fux2023-thermalization,link2024-open,cygorek2024-sublinear}.

Nevertheless, the classic derivations of the path integral formalism
are formulated in the first quantization framework and can not be
directly generalized to the fermion situation. In the second
quantization framework \cite{mahan2000-many,landau1965-quantum}, the
many-particle states are represented in particle number representation
and the many-particle operators are conveniently represented by
annihilation and creation operators. However, a state with a definite
particle number is not the eigenstate of annihilation or creation
operators, which is inconvenient for path integral
formalism. Therefore the coherent states, defined as the eigenstates
of annihilation operators, are introduced. With the aid of coherent
states, the path integral formalism for both boson and fermion systems
can be constructed in a similar spirit.

Due to the existence of the Grassmann variables, the fermion path
integral formalism, unlike the boson one, is usually not directly
evaluated, but serves as an alternative way to derive the Wick's
theorem \cite{negele1998-quantum}. For instance, in dynamical
mean-field theory \cite{georges1996-dynamical} the path integral
formalism is used to formulate the impurity model, but it is still the
perturbative expansion that is used in continuous Monte Carlo impurity
solver \cite{gull2011-continuous}. The situation has changed recently
that some tensor network methods based on path integral formalism
\cite{ng2023-real,thoenniss2023-efficient,chen2024-gtempo,xu2024-grassmann}
are developed. Among them, the Grassmann time-evolving matrix product
operators (GTEMPO) method \cite{chen2024-gtempo,xu2024-grassmann}
employs the TEMPO scheme to evaluate the fermion path integral
expression directly.

The development of tensor network methods for quantum open systems,
such as TEMPO and GTEMPO, relies on the analytical path integral
expression. Nevertheless, the relevant derivation of path integral
expression via coherent states is rather scattered and usually lacks
of details, and this article shall provide a detailed derivation of
path integral formalism for both boson and fermion quantum open
systems using coherent states. The formalism on the imaginary-time
axis, Keldysh contour, and Kadanoff contour are given. The
corresponding generating functional technique, which can be used to
retrieve the environment information from the system correlation
function, is also discussed. In the rest part of this section, we give
a basic introduction to the coherent state. In sections
\ref{sec:boson-path-integral} and \ref{sec:boson-system}, we give a
detailed derivation of boson formalism for partition function and
system correlation functions. Section
\ref{sec:boson-generating-functional} discusses how to use the
generating functional technique to retrieve the environment
information from the system correlation function in boson
formalism. Sections \ref{sec:fermion-path-integral} and
\ref{sec:fermion-generating-functional} discusses the fermion path
integral formalism and the corresponding generating functional
method. A conclusion is given in section
\ref{sec:conclusion}. Throughout this paper, the unit
$\hbar = k_B = 1$ will be used.

\subsection{Boson coherent states}
Here we give a basic introduction to the boson coherent
state. Suppose the number of particles in the single-particle state
$k$ is $n_k$, and the corresponding many-particle state is denoted by
$\ket*{n_k}$. Then the annihilation operator $\hat{b}_k$ and creation
operator $\hat{b}_k^{\dag}$ are defined as
\begin{equation}
  \hat{b}_k \ket*{n_k} = \sqrt{n_k} \ket*{n_k - 1}, \quad \hat{b}_k^{\dag}
  \ket*{n_k} = \sqrt{n_k + 1} \ket*{n_k + 1}.
\end{equation}
The annihilation operator decreases the number of particles in the $k$
state by one, while the creation operator increases it by one, and
thus $\ket*{n_k}$ can not be an eigenstate of $\hat{b}_k$ and
$\hat{b}_k^{\dag}$.  Suppose there are total $p$ orthonormal
single-particle states which are labeled by $1, \ldots, p$. A
many-particle state $\ket*{n_1 \cdots n_p}$ can be generated by
repeating action of the corresponding creation operators on the vacuum
state $\ket*{0}$ that
\begin{equation}
  \ket*{n_1 \cdots n_p} = \frac{(\hat{a}_1^{\dag})^{n_1} \cdots
  (\hat{a}_p^{\dag})^{n_p}}{\sqrt{n_1 ! \cdots n_p !}} \ket*{0}.
\end{equation}
Since the single-particle states are orthonormal, we have the following
orthonormal relation for many-particle states that
\begin{equation}
  \braket*{m_1 \cdots m_p}{n_1 \cdots n_p} = \delta_{m_1 n_1} \cdots
  \delta_{m_p n_p},
\end{equation}
where $\delta_{m n}$ is the Kronecker delta function. The
many-particle identity operator can be written as
\begin{equation}
  \sum_{n_1, \ldots, n_p} \ketbra*{n_1 \cdots n_p}{n_1 \cdots n_p} =1,
  \label{eq:boson-many-particle-identity}
\end{equation}
which leaves any many-particle state unchanged.

A boson coherent state $\ket*{\bm{\varphi}}$ is defined as an
eigenstate of $\hat{b}_k$ for which
\begin{equation}
  \hat{b}_k \ket*{\bm{\varphi}} = \varphi_k \ket*{\bm{\varphi}},
\end{equation}
where $\varphi_k$ is a complex number. Here we directly list some
important properties of the coherent state, and a more detailed
introduction to both boson and fermion coherent states can be found in
the book by Negele and Orland \cite{negele1998-quantum}. A boson
coherent state can be constructed from particle number basis by
\begin{equation}
  \ket*{\bm{\varphi}} = e^{\sum_k \varphi_k \hat{b}_k^{\dag}} \ket*{0} =
  \sum_{n_1, \ldots, n_p} \frac{(\varphi_1\hat{b}_1^{\dag})^{n_1}}{n_1!} \cdots
  \frac{(\varphi_p\hat{b}_p^{\dag})^{n_p}}{n_p!} \ket*{0} =
  \sum_{n_1, \ldots, n_p}\frac{\varphi_1^{n_p}}{\sqrt{n_1!}} \cdots
  \frac{\varphi_p^{n_p}}{\sqrt{n_p!}} \ket*{n_1 \cdots n_p},
\end{equation}
and the corresponding adjoint is
$\bra*{\bm{\varphi}} = \bra*{0} e^{\sum_k \bar{\varphi}_k \hat{b}_k}$,
where $\bar{\varphi}_k$ is the complex conjugate of $\varphi_k$. To
verify that $\ket*{\bm{\varphi}}$ is indeed an eigenstate of
$\hat{b}_k$, we can act $\hat{b}_k$ on it and immediately find that
\begin{equation}
  \hat{b}_k \ket*{\bm{\varphi}} = \varphi_k \qty[ \sum_{n_1, \ldots,n_p}
  \frac{\varphi_1^{n_1}}{\sqrt{n_1!}} \cdots \frac{\varphi^{n_k -1}_k}{\sqrt{(n_k - 1) !}} \cdots
  \frac{\varphi_p^{n_p}}{\sqrt{n_p !}} \ket*{n_1\cdots (n_k - 1) \cdots n_p}] =
  \varphi_k \ket*{\bm{\varphi}}.
\end{equation}
Here we have used the fact that $\hat{b}_k \ket*{n_k = 0} = 0$ and
$n_1, \ldots, n_p$ runs from 0 to $\infty$ in the
summation. Similarly, we have
$\bra*{\bm{\varphi}}\hat{b}_k^{\dag} = \bra*{\bm{\varphi}}
\bar{\varphi}_k$. By definition of $\ket*{\bm{\varphi}}$, the overlap
of two coherent states is given by
\begin{equation}
  \braket*{\bm{\varphi}}{\bm{\varphi}'} = e^{\sum_k\bar{\varphi}_k \varphi_k'}.
\end{equation}
A crucial property of the coherent state is that the many-particle identity
operator \eqref{eq:boson-many-particle-identity} can be expressed by the
coherent state closure relation
\begin{equation}
  \int \mathcal{D}[\bar{\bm{\varphi}} \bm{\varphi}] e^{-\bar{\bm{\varphi}} \bm{\varphi}}
  \ketbra*{\bm{\varphi}}{\bm{\varphi}} = 1,
  \label{eq:boson-coherent-identity}
\end{equation}
where the measure $\mathcal{D}[\bar{\bm{\varphi}} \bm{\varphi}]$ and
the term $e^{-\bar{\bm{\varphi}} \bm{\varphi}}$ are
\begin{equation}
  \mathcal{D} [\bar{\bm{\varphi}} \bm{\varphi}] = \prod_k
  \frac{\dd{\bar{\varphi}_k} \dd{\varphi_k}}{2 \pi i}, \quad
  e^{-\bar{\bm{\varphi}} \bm{\varphi}} = \prod_k e^{-\bar{\varphi}_k \varphi_k} =
  e^{-\sum_k \bar{\varphi}_k\varphi_k} .
\end{equation}
With the aid of this closure relation, the trace of an operator $\hat{A}$ can
be expressed by coherent states as
\begin{equation}
  \Tr\hat{A} = \int \mathcal{D}[\bar{\bm{\varphi}}\bm{\varphi}]
  e^{-\bar{\bm{\varphi}}\bm{\varphi}}
  \mel*{\bm{\varphi}}{\hat{A}}{\bm{\varphi}} .
  \quad \label{eq:boson-coherent-trace}
\end{equation}
There is also a multiple-dimensional Gaussian integral formalism which
is useful in the boson coherent states framework that
\begin{equation}
  \int \mathcal{D} [\bar{\bm{\varphi}} \bm{\varphi}]
  e^{-\sum_{j k} \bar{\varphi}_j S_{j k} \varphi_k + \sum_k (\tilde{J}_k\varphi_k + \bar{\varphi}_k J_k)} =
  [\det S]^{- 1} e^{\sum_{j k}\tilde{J}_j S^{- 1}_{j k} J_k},
  \label{eq:boson-gaussian}
\end{equation}
where $S_{j k}$ is a matrix with a positive Hermitian part. It should
be noted that here $\tilde{J}_k$ need not to be the complex conjugate
of $J_k$.

\subsection{Fermion coherent states}

Because of the anticommutation rule, the fermion coherent states can
not be directly expanded by many-particle states. Due to the
existence of the Pauli exclusion principle, the particle number in a state
$k$ can not exceed one. The annihilation and creation operators are
defined as
\begin{equation}
  \hat{c}_k \ket*{n_k = 0} = 0, \quad
  \hat{c}_k \ket*{n_k = 1} = \ket*{n_k = 0}, \quad
  \hat{c}_k^{\dag} \ket*{n_k = 0} = \ket*{n_k = 1}, \quad
  \hat{c}_k^{\dag} \ket*{n_k = 1} = 0.
\end{equation}
Similar to the boson case, the many-particle identity operator is
\begin{equation}
  \sum_{n_1, \ldots, n_p = 0, 1} \ketbra*{n_1 \cdots n_p}{ n_1 \cdots n_p} = 1.
  \label{eq:fermion-many-particle-identity}
\end{equation}

Unlike the boson operators, fermion operators follow the
anticommutation rule that a minus sign appears when two adjoining
(creation or annihilation) operators interchange. Because of this sign
issue, the eigenstate of $\hat{c}_k$ can not be constructed from
particle number basis directly, and the Grassmann algebra must be
introduced. We associate a Grassmann variable $c_k$ with each
annihilation operator $\hat{c}_k$, and a Grassmann variable
$\bar{c}_k$ with each creation operator $\hat{c}_k^{\dag}$, where
$\bar{c}_k$ is the adjoint of $c_k$. The Grassmann variables
anticommute with each other that
\begin{equation}
  c_j c_k + c_k c_j = c_j \bar{c}_k + \bar{c}_k c_j =
  \bar{c}_jc_k + c_k \bar{c}_j = \bar{c}_j \bar{c}_k + \bar{c}_k\bar{c}_j = 0.
\end{equation}
In particular, we have $c_k c_k = \bar{c}_k \bar{c}_k = 0$. They also
anticommute with each annihilation and creation operator that
\begin{equation}
  c_j \hat{c}_k + \hat{c}_k c_j = c_j \hat{c}_k^{\dag} + \hat{c}_k^{\dag} c_j
  = \bar{c}_j \hat{c}_k + \hat{c}_k \bar{c}_j = \bar{c}_j\hat{c}_k^{\dag} + \hat{c}_k^{\dag} \bar{c}_j = 0.
\end{equation}

The fermion coherent state $\ket*{\bm{c}}$ is then defined with
Grassmann variables as
\begin{equation}
  \ket*{\bm{c}} = e^{-\sum_k c_k \hat{c}_k^{\dag}} \ket*{0} =\prod_k (1 - c_k \hat{c}_k^{\dag}) \ket*{0}.
\end{equation}
Note that (here the property $c_kc_k=0$ is used)
\begin{equation}
  \hat{c}_k (1 - c_k \hat{c}_k^{\dag}) \ket*{0} = c_k \ket*{0} =
  c_k (1 -c_k \hat{c}_k^{\dag}) \ket*{0},
\end{equation}
we have
\begin{equation}
  \hat{c}_k \ket*{\bm{c}} = c_k \prod_j (1 - c_j \hat{c}_j^{\dag}) \ket*{0} =
  \prod_{j \neq k} (1 - c_j \hat{c}_j^{\dag}) c_k (1 - c_k\hat{c}_k^{\dag}) \ket*{0} =
  c_k \prod_j (1 - c_j \hat{c}_j^{\dag}) \ket*{0} = c_k \ket*{\bm{c}}.
\end{equation}
This shows that the coherent state $\ket*{\bm{c}}$ defined in this way
is indeed an eigenstate of annihilation operators, but the eigenvalue
is a Grassmann variable rather than an ordinary number. Similarly, the
adjoint of the coherent state can be written as
$\bra*{\bm{c}} = \bra*{0}e^{- \sum_k \hat{c}_k \bar{c}_k} =
\bra*{0}e^{\sum_k \bar{c}_k \hat{c}_k}$. The overlap of two coherent
states is
\begin{equation}
  \braket*{\bm{c}}{\bm{c}'} = \mel*{0}{\prod_k (1+\bar{c}_k\hat{c}_k)(1-c_k'\hat{c}_k^{\dag})}{0} =
  \prod_k(1 + \bar{c}_k c_k') = e^{\sum_k \bar{c}_k c_k'}.
\end{equation}

The Grassmann variables can be integrated out to yield an ordinary
number. The integration is defined as
\begin{equation}
  \int\dd{c_k} 1 = 0, \quad \int\dd{c_k} c_k = 1, \quad
  \int\dd{\bar{c}}_k 1 = 0, \quad \int \dd{\bar{c}_k}\bar{c}_k = 1.
\end{equation}
Basically, the integration $\int \dd{c_k}$ is nonzero only when acting
it on $c_k$, and the similar rule holds for $\int
\dd{\bar{c}_k}$. Another rule is that the Grassmann integration only
applies to the variable adjacent to it. If the variable is not
adjacent to the integration, it has to be swapped until it is. For
example,
\begin{equation}
  \int\dd{c_k} (\bar{c}_kc_k) = \int\dd{c_k} (-c_k\bar{c}_k) = -\bar{c}_k .
\end{equation}

As in the boson case, the many-particle identity operator
\eqref{eq:fermion-many-particle-identity} can be expressed by the
closure relation that
\begin{equation}
  \int\mathcal{D}[\bar{\bm{c}} \bm{c}] e^{-\bar{\bm{c}} \bm{c}} \ketbra*{\bm{c}}{\bm{c}} = 1,
  \label{eq:fermion-coherent-identity}
\end{equation}
where the measure $\mathcal{D}[\bar{\bm{c}} \bm{c}]$ and the term
$e^{-\bar{\bm{c}} \bm{c}}$ are
\begin{equation}
  \mathcal{D}[\bar{\bm{c}}\bm{c}] = \prod_k\dd{\bar{c}_k}\dd{c_k}, \quad
  e^{-\bar{\bm{c}}\bm{c}}= \prod_k e^{-\bar{c}_kc_k} = e^{-\sum_k\bar{c}_kc_k}.
\end{equation}
With this closure relation, the trace of an operator $\hat{A}$ can be written
as
\begin{equation}
  \Tr\hat{A}=\int\mathcal{D}[\bar{\bm{c}}\bm{c}]e^{-\bar{\bm{c}}\bm{c}}\mel*{-\bm{c}}{\hat{A}}{\bm{c}}.
  \label{eq:fermion-coherent-trace}
\end{equation}
  
In the above expression, the state $\bra*{-\bm{c}}$ has a minus sign
which corresponds to the anticommutation rule of fermion. In Grassmann
algebra, there is also a Gaussian integral formula that
\begin{equation}
  \int\mathcal{D}[\bar{\bm{c}}\bm{c}]
  e^{- \sum_{j k}\bar{c}_j S_{j k} c_k + \sum_k (\tilde{J}_k \bar{a}_k c_k +J_k\bar{c}_k a_k)} =
  [\det S] e^{\sum_j \tilde{J}_j \bar{a}_jS^{-1}_{j k} J_k a_k},
\end{equation}
where $\tilde{J}_j, J_j$ are ordinary numbers and $\bar{a}_j, a_j$ are
Grassmann variables. Unlike the boson case, here $S_{jk}$ need not
to have a positive Hermitian part to ensure the convergence.

\section{Boson Path Integral Formalism}\label{sec:boson-path-integral}

For boson quantum open system, we consider the Caldeira-Leggett type
model
\cite{caldeira1981-influence,caldeira1983-path,caldeira1983-quantum}
where the system is linearly coupled to a continuous boson
environment. The Caldeira-Leggett model is a paradigm to study
quantum open system, whose Hamiltonian can be written as
\begin{equation}
  \hat{H} = \hat{H}_S + \hat{H}_E + \hat{H}_{S E},
\end{equation}
where $\hat{H}_S$ is the system Hamiltonian, $\hat{H}_E$ is the
environment Hamiltonian and $\hat{H}_{S E}$ is the coupling between
the system and the environment. Suppose the eigenstates $\ket*{s}$ of
the system operator $\hat{s}$ forms a complete set of the system with
eigenvalues $s$. The environment consists of a collection of bosons
which couple to the system that
\begin{equation}
  \hat{H}_E = \sum_k \omega_k \hat{b}_k^{\dag} \hat{b}_k, \quad
  \hat{H}_{S E} = \hat{s} \sum_k V_k (\hat{b}_k + \hat{b}_k^{\dag}),
  \label{eq:boson-hamiltonian}
\end{equation}
where $V_k$ is the coupling strength parameter. The environment is
characterized by the spectral function
\begin{equation}
  J (\omega) = \sum_k V_k^2 \delta (\omega - \omega_k).
  \label{eq:boson-spectral-continuous}
\end{equation}

For ease of exposition, we will always consider the situation that the
environment contains only one boson first, where the Hamiltonian
\eqref{eq:boson-hamiltonian} becomes
\begin{equation}
  \hat{H}_E = \omega_0 \hat{b}^{\dag} \hat{b}, \quad
  \hat{H}_{S E} = \hat{s} V(\hat{b} + \hat{b}^{\dag}).
\end{equation}
The single boson case corresponds to the spectral function which
consists of a single delta function that
\begin{equation}
  J(\omega) = V^2\delta(\omega-\omega_0).
  \label{eq:boson-spectral-single}
\end{equation}
The passage from the single boson path integral formalism to the
continuous bosons formalism would be straightforward. We need to only
replace the single boson spectral function
\eqref{eq:boson-spectral-single} by the continuous one
\eqref{eq:boson-spectral-continuous}.

\subsection{On the imaginary-time axis}\label{sec:boson-imaginary}

Here we first consider the imaginary-time formalism, where the
essential ingredients of the path integral formalism will be
illustrated.  The basic idea would remain the same when the formalism
is generalized to the Keldysh and Kadanoff formalisms. The boson
imaginary-time path integral expression has been used to investigate
the phase transition \cite{bray1982-influence}, the quantum mean-force
Gibbs state \cite{chiu2022-numerical}, the correspondence of the
spin-boson model to the Kondo model and the $1/r^2$ Ising model
\cite{weiss1993-quantum}, and so on.

Suppose the whole system is in a thermal equilibrium that the total
density matrix is $\hat{\rho} = e^{-\beta\hat{H}}$, where $\beta$ is
the inverse temperature. Since we are considering the single boson
case, we may define the coherent state as
$\ket*{\varphi} = e^{\varphi \hat{b}^{\dag}}\ket*{0}$.  Then the total
partition function $Z = \Tr e^{- \beta \hat{H}}$ can be expressed by
coherent trace \eqref{eq:boson-coherent-trace} as
\begin{equation}
  Z = \sum_s \int \mathcal{D}[\bar{\varphi}\varphi] e^{-\bar{\varphi} \varphi}
  \mel*{s\varphi}{e^{- \beta \hat{H}}}{s\varphi}.
\end{equation}
Splitting $\beta$ into $M \rightarrow \infty$ pieces that
$\beta = M \delta \tau$, the the above expression becomes
\begin{equation}
  Z = \sum_s \int \mathcal{D}[\bar{\varphi}\varphi] e^{-\bar{\varphi} \varphi}
  \mel*{s\varphi}{e^{- \delta \tau \hat{H}}\cdots e^{- \delta \tau \hat{H}}}{s\varphi}.
\end{equation}
Inserting the whole identity operator
\begin{equation}
  \sum_s \int \mathcal{D}[\bar{\varphi}\varphi] e^{- \bar{\varphi}\varphi}\ketbra*{s \varphi}{s \varphi} 
  \label{eq:boson-whole-identity}
\end{equation}
between every two $e^{- \delta \tau \hat{H}}$ yields
\begin{equation}
  \begin{split}
    Z  =  \sum_{s_0, \ldots, s_{N - 1}} \int\mathcal{D}[\bar{\varphi}_0\varphi_0]
           e^{- \bar{\varphi}_0 \varphi_0} \cdots \int\mathcal{D}[\bar{\varphi}_{M - 1} \varphi_{M - 1}]
           e^{- \bar{\varphi}_{M - 1}\varphi_{M - 1}} 
          \mel*{s_M \varphi_M}{e^{- \delta \tau \hat{H}}}{s_{M - 1}\varphi_{M - 1}} \cdots
           \mel*{s_1 \varphi_1}{e^{- \delta \tau\hat{H}}}{s_0 \varphi_0} 
  \end{split}
  \label{eq:boson-imaginary-partition}
\end{equation}
with the boundary condition
$\bra*{s_M \varphi_M} = \bra*{s_0 \varphi_0}$. This expression states
that the partition function can be obtained via summation over the
functional of all possible paths.

We can employ the first-order Trotter-Suzuki decomposition formula
{\cite{trotter1959-product,suzuki1976-generalized}} to evaluate the
above path integral expression, which states that in the continuous
limit $\delta \tau \rightarrow 0$ we have
\begin{equation}
  e^{- \delta \tau \hat{H}} = e^{- \delta \tau \hat{s} V \hat{b}^{\dag}} e^{-\delta \tau \hat{H}_S}
  e^{- \delta \tau \hat{H}_E} e^{- \delta \tau \hat{s}V \hat{b}}.
\end{equation}
Accordingly,
\begin{equation}
  \mel*{s_{k + 1} \varphi_{k + 1}}{e^{- \delta \tau \hat{H}}}{s_k \varphi_k} =
  \mel*{s_{k + 1}}{e^{- \delta \tau \hat{H}_S}}{s_k}
  e^{h\bar{\varphi}_{k + 1} \varphi_k - \delta \tau V (s_{k + 1}\bar{\varphi}_{k + 1} + s_k \varphi_k)},
\end{equation}
where $h = 1 - \delta \tau \omega_0 = e^{- \delta \tau \omega_0}$. Therefore
the expression \eqref{eq:boson-imaginary-partition} can be written in a form
(with the boundary condition $s_M = s_0$)
\begin{equation}
  Z = Z^{(0)}_E \sum_{\bm{s}} K [\bm{s}] I [\bm{s}], \quad \bm{s} = (s_0, \ldots, s_M),
\end{equation}
where $Z^{(0)}_E = (1 - e^{- \beta \omega_0})^{- 1}$ is free
environment partition function and $K[\bm{s}]$ is the bare system
propagator for which
\begin{equation}
  K [\bm{s}] = \mel*{s_M}{e^{- \delta \tau \hat{H}_S}}{s_{M - 1}} \cdots
  \mel*{s_1}{e^{- \delta \tau \hat{H}_S}}{s_0}.
  \label{eq:boson-imaginary-system-propagator}
\end{equation}

The term $I[\bm{s}]$ is called the influence functional for which
(with the boundary condition $s_M = s_0$ and
$\bar{\varphi}_M = \bar{\varphi}_0$)
\begin{equation}
  I [\bm{s}] = \frac{1}{Z^{(0)}_E} \int \mathcal{D}[\bar{\bm{\varphi}}\bm{\varphi}]
  e^{-\bar{\bm{\varphi}} \bm{\varphi}}
  e^{h \bar{\varphi}_M\varphi_{M - 1} - \delta \tau V (\bar{s}_M \bar{\varphi}_M + s_{M- 1} \varphi_{M - 1})} \cdots
  e^{h \bar{\varphi}_1 \varphi_0 - \delta\tau V (\bar{s}_1 \bar{\varphi}_1 + s_0 \varphi_0)}, \quad
  \label{eq:boson-imaginary-influence-functional}
\end{equation}
where
\begin{equation}
  \mathcal{D} [\bar{\bm{\varphi}}\bm{\varphi}] =
  \prod_{k =0}^{M - 1} \mathcal{D} [\bar{\varphi}_k \varphi_k], \quad
  e^{-\bar{\bm{\varphi}} \bm{\varphi}} = \prod_{k = 0}^{M - 1}e^{- \bar{\varphi}_k \varphi_k} =
  e^{- \sum_{k = 0}^{M - 1}\bar{\varphi}_k \varphi_k} .
\end{equation}
The influence functional
\eqref{eq:boson-imaginary-influence-functional} contains all the
information of the environment's influence on the system, which can be
written in the Gaussian integral form as
\begin{equation}
  I[\bm{s}] = \frac{1}{Z^{(0)}_E} \int \mathcal{D}[\bar{\bm{\varphi}} \bm{\varphi}]
  e^{-\sum_{j,k = 0}^{M-1} \bar{\varphi}_j S_{j k}\varphi_k - \delta\tau V\sum_{k = 0}^{M -1} s_k (\varphi_k+ \bar{\varphi}_k)},
\end{equation}
where the matrix $S_{j k}$ is
\begin{equation}
  S = \mqty[
    1 &  & \cdots &  & - h\\
    - h & 1 &  &  & \\
    & - h & \ddots &  & \vdots\\
    &  & \ddots & 1 & \\
    &  &  & - h & 1].
\end{equation}
Employing the boson Gaussian integral \eqref{eq:boson-gaussian} yields
\begin{equation}
  I [\bm{s}] = \frac{1}{Z^{(0)}_E} [\det S]^{- 1} e^{\delta \tau^2 V^2\sum_{j k} s_j S^{- 1}_{j k} s_k}.
\end{equation}
The determinant of $S$ is simply
\begin{equation}
  \det S = 1 + (- 1)^{M - 1} (- h)^M = 1 - h^M = 1 - e^{- \beta \omega_0},
\end{equation}
and thus $[\det S]^{-1} = (1 - e^{-\beta\omega_0})^{-1}$ just cancels
free environment partition function $Z^{(0)}_E$.

The inverse of matrix $S$ can be evaluated as
\begin{equation}
  S^{- 1} = [1+n (\omega_0)]\mqty[
    1 & h^{M - 1} & h^{M - 2} & \cdots & h^2 & h\\
    h & 1 & h^{M - 1} & \cdots & h^3 & h^2\\
    \vdots & h & \ddots &  & \vdots & \vdots \\
    \vdots & \vdots &  & \ddots & \vdots & \vdots\\
    h^{M - 2} & h^{M - 3} & \cdots & \cdots & 1 & h^{M - 1}\\
    h^{M - 1} & h^{M - 2} & \cdots & \cdots & h & 1],
\end{equation}
where $n (\omega) = (e^{\beta \omega} - 1)^{- 1}$ is the Bose-Einstein
distribution function. The inverse matrix $S^{- 1}$ can be conveniently
represented by the free environment Matsubara Green's function, which is
defined as
\begin{equation}
  D_0 (\omega ; \tau', \tau'') = - \expval*{T \hat{b}_{\omega} (\tau')\hat{b}_{\omega}^{\dag} (\tau'')}_0 =
  \begin{cases}
    -\expval*{\hat{b}_{\omega} (\tau') \hat{b}_{\omega}^{\dag} (\tau'')}_0, & \tau' \geqslant \tau'';\\
    -\expval*{\hat{b}^{\dag}_{\omega} (\tau'') \hat{b}_{\omega} (\tau')}_0, & \tau' < \tau'',
  \end{cases}
\label{eq:boson-imaginary-green}
\end{equation}
where in the free environment we have
\begin{equation}
  \hat{b}_{\omega} (\tau) = e^{\tau \omega \hat{b}^{\dag} \hat{b}} \hat{b}e^{- \tau \omega \hat{b}^{\dag} \hat{b}} =
  \hat{b} e^{- \tau \omega}, \quad
  \hat{b}_{\omega}^{\dag} = e^{\tau \omega \hat{b}^{\dag} \hat{b}}\hat{b}^{\dag} e^{- \tau \omega \hat{b}^{\dag} \hat{b}} =
  \hat{b}^{\dag}e^{\tau \omega}.
\end{equation}
To be explicit,
\begin{equation}
  D_0 (\omega ; \tau', \tau'') = \begin{cases}
    -[1 + n (\omega)] e^{- \omega (\tau' - \tau'')}, & \tau' \geqslant \tau'';\\
    -n (\omega) e^{- \omega (\tau' - \tau'')}, & \tau' < \tau''.
  \end{cases}
\end{equation}

Therefore we have
$S^{-1}_{j k} = -D_0 (\omega_0 ; j \delta \tau, k \delta \tau)$, and
the influence functional can be written as
\begin{equation}
  I[\bm{s}] = e^{-\delta\tau^2V^2\sum_{j k} s_j D_0 (\omega_0; j\delta \tau, k \delta \tau) s_k} .
\end{equation}
In the continuous limit $\delta \tau \rightarrow 0$, the double summation in
the exponent can be recast into a double integral as
\begin{equation}
  I[s (\tau)] = e^{-\int_0^{\beta} \dd{\tau'} \int_0^{\beta} \dd{\tau''}s (\tau') \Lambda (\tau', \tau'') s (\tau'')},
  \label{eq:boson-imaginary-influence-functional-continuous-limit}
\end{equation}
where $s(j \delta \tau) = s_j, s(k \delta \tau) = s_k$. The
correlation function $\Lambda (\tau', \tau'')$ is defined as
\begin{equation}
  \Lambda (\tau', \tau'') = V^2 D_0 (\omega_0 ; \tau', \tau'')=
  \int\dd{\omega}J(\omega)D_0(\omega;\tau',\tau''),
\end{equation}
where $J(\omega)$ is the single boson the spectral function defined in
\eqref{eq:boson-spectral-single}.

Finally, we may write \eqref{eq:boson-imaginary-partition} in the
continuous limit, together with the boundary condition
$s(\tau) = s(0)$, as
\begin{equation}
  Z = Z^{(0)}_E \int \mathcal{D}[s(\tau)] K[s(\tau)] I[s(\tau)],
  \label{eq:boson-imaginary-path-integral-Z}
\end{equation}
where the integral over measure $\mathcal{D}[s(\tau)]$ means summation
over all possible $s(\tau) = (s_M, \ldots, s_0)$ with boundary
condition $s_M = s_0$. Instead of the whole partition function,
sometimes it would be convenient to use the system partition function
which is defined as
\begin{equation}
  Z_S = \frac{Z}{Z^{(0)}_E} = \int\mathcal{D}[s(\tau)] K[s(\tau)] I [s(\tau)].
  \label{eq:boson-imaginary-path-integral}
\end{equation}

Noticing that area integration
$\int_0^{\beta} \dd{\tau'} \int_0^{\beta} \dd{\tau''}$ can be split as
\begin{equation}
  \int_0^{\beta}\dd{\tau'} \int_0^{\beta}\dd{\tau''} =
  \int_{\tau'<\tau''} \dd{\tau'} \dd{\tau''} + \int_{\tau' > \tau''}\dd{\tau'}\dd{\tau''} =
  \int_0^{\beta} \dd{\tau'} \int_0^{\tau'} \dd{\tau''} +\int_0^{\beta} \dd{\tau''}\int_0^{\tau''}\dd{\tau'},
\end{equation}
the influence functional
\eqref{eq:boson-imaginary-influence-functional-continuous-limit} can
be written in the form of
\begin{equation}
  I[s (\tau)] = e^{- \int_0^{\beta} \dd{\tau'} \int_0^{\tau'} \dd{\tau''}s(\tau') \alpha(\tau' - \tau'') s(\tau'')},
  \label{eq:boson-imaginary-influence-functional-auto}
\end{equation}
where $\alpha(\tau' - \tau'')$ is called the autocorrelation function that
\begin{equation}
  \alpha(\tau) = \int\dd{\omega}J(\omega) \{ [1 + n (\omega)] e^{- \tau \omega} + n (\omega) e^{\tau \omega} \}.
  \label{eq:boson-imaginary-auto}
\end{equation}
The autocorrelation function is sometimes written alternatively as
\begin{equation}
  \alpha (\tau) = \int\dd{\omega}J(\omega)\qty[\coth\frac{1}{2}\beta\omega\cosh \tau\omega - \sinh\tau\omega] =
  \int\dd{\omega}J(\omega) \frac{\cosh \qty(\frac{\beta}{2} - \tau)\omega}{\sinh\frac{\beta}{2}\omega}.
\end{equation}

Let us now return to the environment with continuous bosons
\eqref{eq:boson-hamiltonian}. Following the same procedure for the
single boson case, we would obtain the path integral expression for
the system partition function in the same form as
\eqref{eq:boson-imaginary-path-integral} and with the same system
propagator $K[s(\tau)]$ as
\eqref{eq:boson-imaginary-system-propagator}. The only difference is
that the influence functional now becomes
\begin{equation}
  I [s (\tau)] = \prod_k I_k [s (\tau)],
\end{equation}
where
\begin{equation}
  I_k [s (\tau)] = e^{- \int_0^{\beta} \dd{\tau'} \int_0^{\beta} \dd{\tau''} s (\tau') \Lambda_k (\tau', \tau'') s (\tau'')}, \quad
  \Lambda_k(\tau', \tau'') = V_k^2 D_0 (\omega_k; \tau', \tau'').
  \label{eq:boson-imaginary-influence-functional-continuous-k}
\end{equation}
Therefore we can still write the influence functional expression same
to \eqref{eq:boson-imaginary-influence-functional-continuous-limit},
where only the spectral function needs to be modified to the continuous
one \eqref{eq:boson-spectral-continuous} that
\begin{equation}
  \Lambda (\tau', \tau'') = \sum_k V_k^2 D_0 (\omega_k ; \tau', \tau'') =
  \int\dd{\omega} J (\omega) D_0 (\omega ; \tau', \tau''),\quad
  J(\omega)=\sum_kV_k^2\delta(\omega-\omega_k).
  \label{eq:boson-correlation-continuous}
\end{equation}

\subsection{On the Keldysh contour}\label{sec:boson-keldysh}

Now let us consider the real-time dynamics of the Caldeira-Leggett
model in a nonequilibrium setup, which is widely used in the QuAPI and
TEMPO scheme. Suppose that the density matrix of the whole system is
in a product form at the initial time $t = 0$ for which
\begin{equation}
  \hat{\rho}(0) = \hat{\rho}_S(0) \otimes \hat{\rho}_E,
\end{equation}
where $\hat{\rho}_S(0)$ is the initial system density matrix and
$\hat{\rho}_E$ is the initial environment density matrix. It is
assumed that at the initial time the environment is in thermal
equilibrium that $\hat{\rho}_E = e^{-\beta\hat{H}_E}$. The evolution
of $\hat{\rho} (t)$ follows the von Neumann equation, which states
that at the final time $t_f$ the whole density matrix is
\begin{equation}
  \hat{\rho} (t_f) = e^{- i \hat{H} t_f} \hat{\rho} (0) e^{i \hat{H} t_f} .
\end{equation}
Splitting $t_f$ into $N \rightarrow \infty$ pieces that $t_f = N \delta t$,
the partition function at time $t_f$ can be written as
\begin{equation}
  Z (t_f) = \Tr\relax[e^{- i \hat{H} t_f} \hat{\rho} (0) e^{i \hat{H} t_f}] =
  \Tr\relax[e^{- i \hat{H} \delta t} \cdots e^{- i \hat{H} \delta t}\hat{\rho} (0) e^{i \hat{H} \delta t}
  \cdots e^{i \hat{H} \delta t}].
  \label{eq:boson-keldysh-partition-split}
\end{equation}
We can insert the identity operator \eqref{eq:boson-whole-identity} between
every two exponents and obtain that
\begin{equation}
  \begin{split}
    Z (t_f) = & \sum_{\bm{s}} \int \mathcal{D}[\bar{\bm{\varphi}}\bm{\varphi}]e^{-\bar{\bm{\varphi}} \bm{\varphi}}
                \mel*{s \varphi}{e^{- i\hat{H} \delta t}}{s_{N - 1}^+ \varphi_{N - 1}^+} \cdots
                \mel*{s_1^+ \varphi_1^+}{e^{- i \hat{H} \delta t}}{s_0^+ \varphi_0^+}\\
              & \times \mel*{s_0^+ \varphi_0^+}{\hat{\rho}(0)}{s_0^- \varphi_0^-}
                \mel*{s_0^-\varphi_0^-}{e^{i \hat{H} \delta t}}{s_1^-\varphi_1^-} \cdots
                \mel*{s_{N - 1}^- \varphi_{N - 1}^-}{e^{i\hat{H} \delta t}}{s \varphi},
  \end{split}
  \label{eq:boson-keldysh-partition-unrelabel}
\end{equation}
where
\begin{equation}
  \bm{s} = (s_0^{\pm}, \ldots, s_{N - 1}^{\pm}, s),
\end{equation}
and
\begin{equation}
  \int \mathcal{D}[\bar{\bm{\varphi}} \bm{\varphi}] e^{-\bar{\bm{\varphi}} \bm{\varphi}} =
  \int \mathcal{D}[\bar{\varphi}_0^{\pm}\varphi_0^{\pm}] e^{- \bar{\varphi}_0^{\pm}\varphi_0^{\pm}} \cdots
  \int \mathcal{D}[\bar{\varphi}_{N - 1}^{\pm}\varphi_{N - 1}^{\pm}] e^{- \bar{\varphi}_{N - 1}^{\pm} \varphi_{N -1}^{\pm}}
  \int \mathcal{D}[\bar{\varphi}\varphi] e^{-\bar{\varphi} \varphi} .
\end{equation}
At the both ends of \eqref{eq:boson-keldysh-partition-unrelabel},
there is no superscript and subscript in $\bra*{s \varphi}$ and
$\ket*{s \varphi}$, and we relabel them as $\bra*{s^+_N \varphi_N^+}$
and $\ket*{s_N^- \varphi_N^-}$ separately. After this relabeling,
every variable is assigned with a time step, and then we rewrite
\eqref{eq:boson-keldysh-partition-unrelabel} as
\begin{equation}
  \begin{split}
    Z (t_f) = & \sum_{\bm{s}} \int \mathcal{D}[\bar{\bm{\varphi}} \bm{\varphi}] e^{-\bar{\bm{\varphi}} \bm{\varphi}}
                \mel*{s^+_N \varphi_N^+}{e^{- i \hat{H} \delta t}}{s_{N - 1}^+ \varphi_{N - 1}^+} \cdots
                \mel*{s_1^+ \varphi_1^+}{e^{- i \hat{H} \delta t}}{s_0^+ \varphi_0^+}\\
              & \times \mel*{s_0^+\varphi_0^+}{\hat{\rho} (0)}{s_0^- \varphi_0^-}
                \mel*{s_0^-\varphi_0^-}{e^{i \hat{H} \delta t}}{s_1^-\varphi_1^-} \cdots
                \mel*{s_{N - 1}^- \varphi_{N - 1}^-}{e^{i\hat{H} \delta t}}{s^-_N \varphi^-_N}
  \end{split}
  \label{eq:boson-keldysh-partition}
\end{equation}
with the boundary condition
$\bra*{s_N^+ \varphi_N^+} = \bra*{s\varphi}$ and
$\ket*{s_N^- \varphi^-_N}=\ket*{s\varphi}$.

Similar to the imaginary time case, we employ the Trotter-Suzuki
decomposition in the way that
\begin{equation}
  e^{\mp i \hat{H} \delta t} = e^{\mp i \delta t V \hat{s} \hat{b}^{\dag}}
  e^{\mp i \hat{H}_S \delta t} e^{\mp i \hat{H}_E \delta t} e^{\mp i \delta tV \hat{s} \hat{b}} .
\end{equation}
Accordingly, we have
\begin{equation}
  \mel*{s_{k + 1}^+ \varphi_{k + 1}^+}{e^{- i \hat{H} \delta t}}{s_k^+\varphi_k^+} =
  \mel*{s_{k + 1}^+}{e^{- i \hat{H}_S \delta t}}{s_k^+}
  e^{g \bar{\varphi}_{k + 1}^+ \varphi_k^+ -i \delta t V(s_{k + 1}^+ \bar{\varphi}^+_{k + 1} + s_k^+ \varphi^+_k)},
\end{equation}
and
\begin{equation}
  \mel*{s_k^- \varphi_k^-}{e^{i \hat{H} \delta t}}{s_{k + 1}^- \varphi_{k +1}^-} =
  \mel*{s_k^-}{e^{i \hat{H}_S \delta t}}{s_{k + 1}^-}
  e^{\bar{g}  \bar{\varphi}_k^- \varphi_{k + 1}^- + i \delta t V(s_k^- \bar{\varphi}_k^- + s^-_{k + 1} \varphi_{k + 1}^-)},
\end{equation}
where $g = 1 - i \omega_0 \delta t = e^{- i \omega_0 \delta t}$ and
$\bar{g} = 1 + i \omega_0 \delta t = e^{i \omega_0 \delta t}$. Since
at the initial time the whole system is in a product state that
$\hat{\rho}(0) = \hat{\rho}_S (0) \otimes \hat{\rho}_E$, the
expression \eqref{eq:boson-keldysh-partition} can be written as
\begin{equation}
  Z (t_f) = Z^{(0)}_E \sum_{\bm{s}} K[\bm{s}] I[\bm{s}],
\end{equation}
where $K[\bm{s}]$ is the bare system propagator (with boundary
condition $s_N^+ = s_N^-=s$)
\begin{equation}
  K[\bm{s}] = \mel*{s_N^+}{e^{- i \hat{H}_S \delta t}}{s_{N - 1}^+} \cdots
  \mel*{s_0^+}{\hat{\rho}_S (0)}{s_0^-} \cdots
  \mel*{s_{N - 1}^-}{e^{i \hat{H}_S \delta t}}{s_N^-},
\end{equation}
and the influence functional $I [\bm{s}]$ is (with boundary condition
$s_N^+ = s_N^-=s,\bar{\varphi}_N^+=\bar{\varphi},\varphi_N^-=\varphi$)
\begin{equation}
  \begin{split}
    I [\bm{s}] = & \frac{1}{Z^{(0)}_E} \int \mathcal{D}
                   [\bar{\bm{\varphi}} \bm{\varphi}] e^{-\bar{\bm{\varphi}} \bm{\varphi}}
                   e^{g\bar{\varphi}_N^+ \varphi_{N - 1}^+ - i \delta t V (s_N^+\bar{\varphi}_N^+ + s_{N - 1}^+ \varphi_{N - 1}^+)} \cdots
                   e^{g\bar{\varphi}_1^+ \varphi_0^+ - i \delta t V (s_1^+\bar{\varphi}_1^+ + s_0^+ \varphi_0^+)} \\
                 & \times \mel*{\varphi_0^+}{\hat{\rho}_E}{\varphi_0^-}
                   e^{\bar{g} \bar{\varphi}_0^- \varphi_1^- +i\delta t V (s_0^-\bar{\varphi}^-_0 + s_1^- \varphi_1^-)} \cdots
                   e^{\bar{g}\bar{\varphi}_{N - 1}^- \varphi_N^- +i\delta t V ( s_{N - 1}^-\bar{\varphi}_{N - 1}^- + s_N^- {\varphi_N^-})} .
  \end{split}
  \label{eq:boson-keldysh-influence-functional}
\end{equation}

Let us first evaluate the term
$\mel*{\varphi_0^+}{\hat{\rho}_E}{\varphi_0^-}$. Recall that
$\hat{\rho}_E = e^{- \beta \hat{H}_E}$ and split
$\beta = M \delta \tau$ with $M \rightarrow \infty$, then
\begin{equation}
  \mel*{\varphi_0^+}{\hat{\rho}_E}{\varphi_0^-} = 
  \mel*{\varphi_0^+}{e^{- \delta \tau \hat{H}_E} \cdots e^{- \delta \tau \hat{H}_E}}{\varphi_0^-}.
\end{equation}
Inserting the identity operator \eqref{eq:boson-coherent-identity} between
every two exponents yields an imaginary time path integral expression that
\begin{equation}
  \mel*{\varphi_0^+}{\hat{\rho}_E}{\varphi_0^-} =
  \int \mathcal{D}[\bar{\varphi}_1^{\sim}\varphi^{\sim}_1] \cdots
  \int\mathcal{D}[\bar{\varphi}_{M-1}^{\sim} \varphi^{\sim}_{M-1}] e^{- \sum_{k =1}^{M-1} \bar{\varphi}_k^{\sim} \varphi^{\sim}_k}
  e^{h\bar{\varphi}_0^+ \varphi^{\sim}_{M-1}} e^{h \bar{\varphi}_{M-1}^{\sim} \varphi^{\sim}_{M-2}} \cdots
  e^{h \bar{\varphi}_2^{\sim}\varphi^{\sim}_1} e^{h \bar{\varphi}^{\sim}_1 \varphi_0^+},
\end{equation}
where $h = 1-\delta\tau\omega_0 = e^{-\delta\tau\omega_0}$. This
expression can be written in Gaussian integral form as
\begin{equation}
  \mel*{\varphi_0^+}{\hat{\rho}_E}{\varphi_0^-} =
  \int \mathcal{D}[\bar{\varphi}^{\sim}_1 \varphi^{\sim}_1] \cdots
  \int \mathcal{D}[\bar{\varphi}_{M - 1}^{\sim} \varphi^{\sim}_{M - 1}]
  e^{- \sum_{j, k =1}^{M - 1} \bar{\varphi}^{\sim}_j S_{j k} \varphi^{\sim}_k + h\bar{\varphi}_0^+ \varphi^{\sim}_{M - 1} + h\bar{\varphi}^{\sim}_1 \varphi_0^-},
\end{equation}
where $S$ and its inverse $S^{- 1}$ are
\begin{equation}
  S = \mqty[
  1 &  &  &  &  & \\
  - h & 1 &  &  &  & \\
  & - h & \ddots &  &  & \\
  &  &  & \ddots &  & \\
  &  &  & - h & 1 & \\
  &  &  &  & - h & 1], \quad
  S^{-1} = \mqty[
  1 &  &  &  &  & \\
  h & 1 &  &  &  & \\
  h^2 & h & \ddots &  &  & \\
  \vdots & \vdots & \ddots & \ddots &  & \\
  h^{M - 3} & h^{M - 4} & \cdots & h & 1 & \\
  h^{M - 2} & h^{M - 3} & \cdots & \cdots & h & 1].
\end{equation}
Then employing the Gaussian integral formula \eqref{eq:boson-gaussian}
we obtain that
\begin{equation}
  \mel*{\varphi_0^+}{\hat{\rho}_E}{\varphi_0^-} =
  e^{h^2\bar{\varphi}_0^+ S^{- 1}_{M-1, 1} \varphi_0^-} =
  e^{e^{-\beta\omega_0} \bar{\varphi}_0^+ \varphi_0^-} .
\end{equation}

\begin{figure}[htbp]
  \centerline{\includegraphics[]{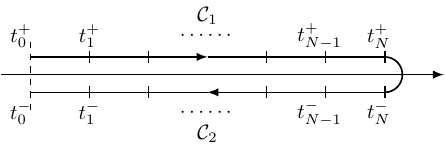}}
  \caption{The Keldysh contour
    $\mathcal{C} = \mathcal{C}_1 \cup \mathcal{C}_2$, where the upper
    part is the forward (1st) branch and lower part is the backward
    (2nd) branch.}
  \label{fig:keldysh}
\end{figure}

To evaluate the rest part of the influence functional, it would be
convenient to align the time steps in the order of
$(t_0^+, \ldots, t_N^+ = t_N^-, \ldots, t_0^-)$, then the set of time
steps forms a closed time contour $\mathcal{C}$ shown in figure
\ref{fig:keldysh}. It should be noted that $t_N^+$ and $t_N^-$ are in
fact the same time step according to the boundary condition. Such a
contour is usually referred to as the Keldysh contour
{\cite{keldysh1965-diagram,lifshitz1981-physical,kamenev2009-keldysh,wang2013-nonequilibrium}}.
The path from $t_0^+$ to $t_N^+$ corresponds to the forward evolution
operator $e^{- i \hat{H} t_f}$ and thus we call it the forward
branch. Similarly, the path from $t_N^-$ to $t_0^-$ is called the
backward branch. We denote the forward branch as the 1st branch
$\mathcal{C}_1$ and the backward branch as the 2nd branch
$\mathcal{C}_2$.

Now we define the contour time interval $\dd{t}$ for which
$\dd{t} = \delta t$ if it is on the forward branch and
$\dd{t} = -\delta t$ on the backward branch. Denote
$\bm{s} = (s_0^+,s_1^+,\ldots,s^+_N = s_N^-,\ldots,s_1^-,s_0^-)$,
the influence functional \eqref{eq:boson-keldysh-influence-functional}
can be written as
\begin{equation}
  I[\bm{s}] = \frac{1}{Z^{(0)}_E} \int \mathcal{D}[\bar{\bm{\varphi}} \bm{\varphi}]
  e^{- \sum_{j, k = 0}^{2N} \bar{\varphi}_j S_{j k} \varphi_k - i \dd t V \sum_{k = 0}^{2 N}s_k (\varphi_k + \bar{\varphi}_k)},
\end{equation}
where the matrix $S$ is
\begin{equation}
  S = \mqty[
  1 &  &  &  &  &  & - e^{- \beta \omega_0}\\
  - g & 1 &  &  &  &  & \\
  & - g & \ddots &  &  &  & \\
  &  & \ddots & 1 &  &  & \\
  &  &  & - \bar{g} & \ddots &  & \\
  &  &  &  & \ddots & 1 & \\
  &  &  &  &  & - \bar{g} & 1].
\end{equation}
The determinant of $S$ is
\begin{equation}
  \det S = 1 + (- 1)^{2 N} (- g)^N (- \bar{g})^N (-e^{-\beta\omega_0}) = 1 - e^{-\beta\omega_0},
\end{equation}
and thus $[\det S]^{- 1}$ cancels $Z^{(0)}_E$. The inverse of $S$ is
[here $n = (e^{\beta\omega_0} - 1)^{-1}$]
\begin{equation}
  S^{- 1} = \mqty[
  1 + n & n \bar{g} & \cdots & n \bar{g}^N & \cdots & n\bar{g} & n\\
  (1 + n) g & 1 + n &  & \vdots &  & n & n g\\
  \vdots & \vdots & \ddots & n \bar{g} &  & \vdots & \vdots\\
  (1 + n) g^N & (1 + n) g^{N - 1} & \cdots & 1 + n &  & n g^{N - 1} & ng^N\\
  \vdots & \vdots &  & (1 + n) \bar{g} & \ddots  & \vdots & \vdots\\
  (1 + n) g & 1 + n &  & \vdots &   & 1 + n & n g\\
  1 + n & (1 + n) \bar{g} & \cdots & (1 + n) \bar{g}^N & \cdots & (1+n) \bar{g} & 1+n].
\end{equation}
As in the imaginary-time case, the inverse matrix $S^{- 1}$ can be
conveniently represented by the Green's function. The free environment Green's
function on the Keldysh contour is defined as
\begin{equation}
  D_0 (\omega ; t', t'') = - i \expval*{T_{\mathcal{C}} \hat{b}_{\omega} (t')\hat{b}_{\omega}^{\dag} (t'')}_0 =
  \begin{cases}
    - i\expval*{\hat{b}_{\omega} (t') \hat{b}_{\omega}^{\dag} (t'')}_0,& t' \succcurlyeq t'';\\
    - i\expval*{\hat{b}^{\dag}_{\omega} (t'') \hat{b}_{\omega} (t')}_0,& t' \prec t'',
  \end{cases}
  \label{eq:boson-keldysh-green}
\end{equation}
where in the free environment
\begin{equation}
  \hat{b}_{\omega}(t) = e^{i\omega\hat{b}^{\dag} \hat{b} t}\hat{b}e^{-i\omega \hat{b}^{\dag}\hat{b}t} =
  \hat{b} e^{-i\omega t}, \quad
  \hat{b}_{\omega}^{\dag}(t) = e^{i \omega \hat{b}^{\dag} \hat{b} t}\hat{b}^{\dag} e^{- i \omega \hat{b}^{\dag} \hat{b} t} =
  \hat{b}^{\dag} e^{i\omega t}.
\end{equation}
Here $t' \succ t''$ means that $t'$ suceeds $t''$ on the contour, and $T_{\mathcal{C}}$ is
the contour order operator. Depending on which branch the arguments are on,
the contour-ordered Green's function can be split into four blocks as
\begin{equation}
  D_0 (\omega ; t', t'') = \mqty[
    D^{11}_0 (\omega ; t', t'') & D^{12}_0 (\omega ; t', t'')\\
    D_0^{21} (\omega ; t', t'') & D_0^{22} (\omega ; t', t'')].
\end{equation}
In $D_0^{11} (\omega ; t', t'')$, both $t', t''$ are on\,the 1st
(forward) branch. This means that if $t' \succ t''$ on the contour we
have $t' > t''$on the normal time axis, and vice versa. Accordingly,
\begin{equation}
  D^{11}_0 (\omega ; t', t'') = - i \expval*{T \hat{b}_{\omega} (t')\hat{b}^{\dag}_{\omega} (t'')}_0 =
  \begin{cases}
    - i [1 + n (\omega)] e^{- i \omega (t' - t'')}, & t' \geqslant t'' ;\\
    - i n (\omega) e^{- i \omega (t' - t'')}, & t' < t'',
  \end{cases}
  \label{eq:boson-keldysh-green-pp}
\end{equation}
where $T$ is the normal time ordering operator and
$n(\omega) = (e^{\beta\omega} - 1)^{- 1}$ is the Bose-Einstein
distribution function. On the contrary, since in $D^{22}_0$ both
$t', t''$ are on the 2nd (backward) branch, $t' \succ t''$ on the
contour means $t' < t''$ on the normal time axis, and vice versa.
Therefore on the time axis $D^{22}_0$ is the anti time-ordered Green's
function for which
\begin{equation}
  D^{22}_0 (\omega ; t', t'') = - i \expval*{\bar{T} \hat{b}_{\omega} (t')\hat{b}_{\omega}^{\dag} (t'')}_0 =
  \begin{cases}
    -i n(\omega) e^{- i \omega (t' - t'')}, & t' \geqslant t'';\\
    -i [1 + n (\omega)] e^{- i \omega (t' - t'')}, & t' < t''.
  \end{cases}
  \label{eq:boson-keldysh-green-mm}
\end{equation}
In $D^{12}_0$, $t'$ is on the 1st branch and $t''$ is on the 2nd branch, while
the situation is just the opposite in $D^{21}_0$. This means that $t''$ always
succeeds $t'$ in $D^{12}_0$, and $t'$ always succeeds $t''$ in $D^{21}_0$,
therefore we have
\begin{equation}
  D^{12}_0 (\omega ; t', t'') = - i \expval*{\hat{b}_{\omega}^{\dag} (t'')\hat{b}_{\omega} (t')}_0 =
  -in(\omega)e^{- i \omega (t' - t'')},
  \label{eq:boson-keldysh-green-pm}
\end{equation}
and
\begin{equation}
  D^{21}_0 (\omega ; t', t'') = -i \expval*{\hat{b}_{\omega} (t')\hat{b}^{\dag}_{\omega} (t'')}_0 =
  -i[1 + n (\omega)] e^{- i \omega(t' - t'')}.
  \label{eq:boson-keldysh-green-mp}
\end{equation}

It can be seen that $S^{- 1}$ can be expressed by the contour ordered Green's
function as
\begin{equation}
  S^{- 1} = i \mqty[
    D^{11}_0 & D^{12}_0\\
    D^{21}_0 & D^{22}_0],
\end{equation}
and in the continuous limit $\delta t \rightarrow 0$ we can write the
influence functional \eqref{eq:boson-keldysh-influence-functional} as
\begin{equation}
  I [s(t)] = e^{-\int_{\mathcal{C}} \dd{t'} \int_{\mathcal{C}}\dd{t''}s (t') \Lambda (t', t'')s (t'')},
  \label{eq:boson-keldysh-influence-functional-continuous-limit}
\end{equation}
where
\begin{equation}
  \Lambda (t', t'') = i V^2 D_0 (\omega_0 ; t', t'')=
  i\int\dd{\omega}J(\omega)D_{0}(\omega;t',t'').
\end{equation}
Note that the sign of $\dd t'$ and $\dd t''$ depends on the branch of
the contour.

The correlation function $\Lambda (t', t'')$ is linear to the contour
ordered Green's function $D_0 (\omega ; t', t'')$, and it can be also
split into four blocks on the normal time axis as
\begin{equation}
  \Lambda (t', t'') = \mqty[
    \Lambda^{11} (t', t'') & \Lambda^{12} (t', t'')\\
    \Lambda^{21} (t', t'') & \Lambda^{22} (t', t'')].
\end{equation}
Similar to the imaginary-time situation, employing the specific
expressions
\eqref{eq:boson-keldysh-green-pp}-\eqref{eq:boson-keldysh-green-mp}
and splitting the area integration
$\int_0^{t_f}\dd{t'}\int_0^{t_f}\dd{t''}$ as
$\int_{t'>t''}\dd{t'}\dd{t''}+\int_{t'<t''}\dd{t'}\dd{t''}$ the
influence functional
\eqref{eq:boson-keldysh-influence-functional-continuous-limit} can be
written on the normal time axis as
\begin{equation}
  I [s^{\pm}(t)] = e^{-\int_0^{t_f} \dd{t'} \int_0^{t'} \dd{t''} [s^+(t') - s^-(t')][\alpha(t' - t'')s^+(t'') - \bar{\alpha}(t' - t'')s^-(t'')]},
\end{equation}
where $\alpha (t)$ is the autocorrelation function and
$\bar{\alpha} (t)$ is its complex conjugate for which
\begin{equation}
  \alpha (t) = \int\dd{\omega}J(\omega) \{[1 + n(\omega)] e^{- i \omega t} + n(\omega) e^{i \omega t}]\}.
  \label{eq:boson-keldysh-auto}
\end{equation}
Alternatively, the autocorrelation function can be also written as
\begin{equation}
  \alpha(t) = \int\dd{\omega}J(\omega)\qty[\coth\frac{1}{2} \beta \omega \cos \omega t - i\sin \omega t] =
  \int\dd{\omega}J(\omega)\frac{\cosh \qty( \frac{1}{2} \beta - it) \omega}{\sinh\frac{1}{2}\beta\omega} .
\end{equation}

The Keldysh formalism can be generalized to the situation where the
environment consists of multiple baths in a straightforward manner. In
this case, the environment Hamiltonian can be written as
$\hat{H}_E = \sum_{\alpha} \hat{H}_{\alpha}$, where $\hat{H}_{\alpha}$
is the Hamiltonian of $\alpha$th bath coupled to the system via
$\hat{H}_{S\alpha}$ as
\begin{equation}
  \hat{H}_{\alpha} = \omega_{\alpha} \hat{b}^{\dag}_{\alpha} \hat{b}_{\alpha},\quad
  \hat{H}_{S \alpha} = \hat{s} \sum_{\alpha} V_{\alpha} (\hat{b}_{\alpha} +\hat{b}_{\alpha}^{\dag}) .
\end{equation}
The spectral function for $\alpha$th bath is
$J_{\alpha}(\omega)=V^2_{\alpha}\delta(\omega-\omega_{\alpha})$. The
baths are assumed to be in thermal equilibrium with their own
temperature respectively at the initial time. The corresponding density
matrix of $\alpha$th bath at the initial time is
$\hat{\rho}_{\alpha} = e^{- \beta_{\alpha}\hat{H}_{\alpha}}$, where
$\beta_{\alpha}$ is corresponding inverse temperature. In continuous
bosons case, the Hamiltonian of $\alpha$th bath and the coupling are
\begin{equation}
  \hat{H}_{\alpha} = \sum_k \omega_{\alpha k} \hat{b}_{\alpha k}^{\dag}\hat{b}_{\alpha k}, \quad
  \hat{H}_{S \alpha} = \hat{s} \sum_k V_{\alpha k}(\hat{b}_{\alpha k} + \hat{b}_{\alpha k}^{\dag}),
\end{equation}
and each bath is associated with a spectral function
$J_{\alpha}(\omega) =\sum_kV_{\alpha k}^2 \delta(\omega-\omega_{\alpha k})$.

Here we only consider the situation that all baths coupled to the
system via a same system operator $\hat{s}$. When different bath
couples to the system via different system operators the situation
becomes more complicated, see
Refs. \cite{gribben2022-exact,chen2024-path} and the references
therein. Following the same procedure, we would obtain the influence
functional of the same form as
\eqref{eq:boson-keldysh-influence-functional-continuous-limit}, where
we only need to replace the correlation function by
\begin{equation}
  \Lambda (t', t'') = \sum_{\alpha} \Lambda_{\alpha}(t', t''),
\end{equation}
where $\Lambda_{\alpha} (t', t'')$ is the correlation function due to
the $\alpha$th bath that
\begin{equation}
  \Lambda_{\alpha} (t', t'') = i \int \dd{\omega} J_{\alpha}(\omega)D_{\alpha} (\omega ; t', t'').
\end{equation}
Here ${D_{\alpha}} $ is the free bath contour ordered Green's function
for the $\alpha$th bath.

\subsection{On the Kadanoff contour}\label{sec:boson-kadanoff}

\begin{figure}[htbp]
  \centerline{\includegraphics[]{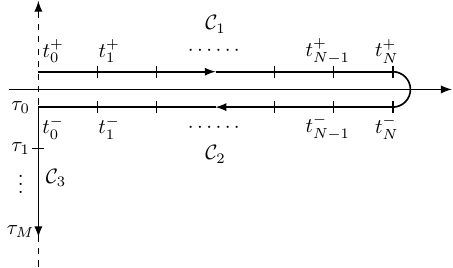}}
  \caption{The L-shaped Kadanoff-Baym contour
    $\mathcal{C} = \mathcal{C}_1 \cup \mathcal{C}_2 \cup
    \mathcal{C}_3$, where the arrows indicate the ordering of the
    contour.}
  \label{fig:kadanoff}
\end{figure}

The Keldysh formalism starts from an initial state with a decoupled
system and environment, which is suitable to describe the
nonequilibrium dynamics. Let us now consider the situation that at the
initial time, the system and environment are in thermal equilibrium as
a whole, which can be used to evaluate the equilibrium correlation
functions \cite{shao2002-iterative,chen2024-solving}. The whole
density matrix at the initial time is
$\hat{\rho} (0) = e^{- \beta \hat{H}}$, then at time $t_f$ the density
matrix becomes
\begin{equation}
  \hat{\rho}(t_f) = e^{-i\hat{H} t_f} e^{-\beta \hat{H}} e^{i \hat{H} t_f}.
\end{equation}
The partition function is
$Z (\beta, t_f) = \Tr\relax[e^{- i \hat{H} t_f} e^{- \beta \hat{H}}
e^{i\hat{H} t_f}] = \Tr\relax [e^{- \beta \hat{H}} e^{i\hat{H}t_f}
e^{-i \hat{H} t_f}]$, where we have used the cyclic property of the
trace. If reading the terms inside the square bracket from right to
left, we may regard the evolution starts from 0 to $t_f$ by a forward
evolution $e^{-i\hat{H} t_f}$, then returns back to initial time by a
backward evolution $e^{i \hat{H} t_f}$, and finally goes to an
imaginary time $- i \beta$ by an imaginary evolution
$e^{- \beta \hat{H}}$. The evolution goes along a L-shaped contour
$\mathcal{C}$, which is usually referred to as the Kadanoff-Baym
contour \cite{kadanoff1962-quantum,aoki2014-nonequilibrium}.  The
contour consists of three branches: the forward (1st) branch
$\mathcal{C}_1$, the backward (2nd) branch $\mathcal{C}_2$ and the
imaginary (3rd) branch $\mathcal{C}_3$, as shown in figure
\ref{fig:kadanoff}.

Splitting $t_f = N \delta t, \beta = M \delta \tau$ with $N \rightarrow \infty,
M \rightarrow \infty$, the partition function can be written as
\begin{equation}
  \begin{split}
    Z(\beta, t_f)  = & \sum_{\bm{s}} \int \mathcal{D}[\bar{\bm{\varphi}} \bm{\varphi}]
                       e^{-\bar{\bm{\varphi}}\bm{\varphi}}
                       \mel*{s_M^{\sim}\varphi^{\sim}_M}{e^{- \delta \tau \hat{H}}}{s^{\sim}_{M-1}\varphi^{\sim}_{M-1}} \cdots
                       \mel*{s^{\sim}_1\varphi^{\sim}_1}{e^{- \delta \tau \hat{H}}}{s^{\sim}_0\varphi^{\sim}_0} \\
                     & \times \mel*{s_0^-\varphi_0^-}{e^{i \hat{H} \delta t}}{s_1^-\varphi_1^-} \cdots
                       \mel*{s_{N - 1}^-\varphi_{N - 1}^-}{e^{i\hat{H} \delta t}}{s_N^- \varphi_N^-} \\
                     & \times \mel*{s_N^+\varphi_N^+}{e^{- i \hat{H} \delta t}}{s_{N -1}^+ \varphi_{N - 1}^+} \cdots
                       \mel*{s_1^+\varphi_1^+}{e^{- i\hat{H} \delta t}}{s_0^+\varphi_0^+}, 
  \end{split}
\end{equation}
where the boundary condition is
$\bra*{s_M^{\sim} \varphi_M^{\sim}} = \bra*{s_0^+ \varphi_0^+},
\bra*{s_0^-\varphi_0^-} = \bra*{s_0^{\sim} \varphi_0^{\sim}},
\bra*{s_N^+\varphi_N^+} = \bra*{s_N^- \varphi_N^-}$. Following the
same procedure described in previous subsections, we have
\begin{equation}
  Z(\beta, t_f) = Z^{(0)}_E \int\mathcal{D}[s(t)] K[s(t)] I[s(t)],
\end{equation}
where the system propagator is
\begin{equation}
  \begin{split}
    K[s(t)] = & \mel*{s_M^{\sim}}{e^{- \delta \tau \hat{H}_S}}{s_{M -1}^{\sim}} \cdots
                \mel*{s_1^{\sim}}{e^{- \delta \tau \hat{H}_S}}{s_0^{\sim}}
                \mel*{s_0^-}{e^{i \hat{H}_S \delta t}}{s_1^-}\cdots\\
              & \times \mel*{s_{N - 1}^-}{e^{i \hat{H}_S \delta t}}{s_N^-}
                \mel*{s_N^+}{e^{- i \hat{H}_S \delta t}}{s_{N - 1}^+} \cdots
                \mel*{s_1^+}{e^{- i \hat{H}_S \delta t}}{s_0^+}.
  \end{split}
\end{equation}
The influence functional is
\begin{equation}
  \begin{split}
    I [s (t)] = & \frac{1}{Z^{(0)}_E}\int\mathcal{D}[\bar{\bm{\varphi}}\bm{\varphi}]e^{-\bar{\bm{\varphi}}\bm{\varphi}}
                  e^{h\bar{\varphi}_M^{\sim}\varphi_{M-1}^{\sim} - \delta\tau V(s_M^{\sim}\bar{\varphi}_M^{\sim} + s_{M-1}^{\sim} \varphi_{M - 1}^{\sim})} \cdots
                  e^{h \bar{\varphi}_1^{\sim} \varphi_0^{\sim} -\delta \tau V (s_1^{\sim} \bar{\varphi}_1^{\sim} + s_0^{\sim}\varphi_0^{\sim})}\\
                & \times e^{\bar{g} \bar{\varphi}_0^- \varphi_1^- + i \delta t V(s_0^- \bar{\varphi}_0^- + s_1^- \varphi_1^-)} \cdots
                  e^{\bar{g}\bar{\varphi}_{N - 1}^- \varphi_N^- + i \delta t V (s_{N - 1}^-\bar{\varphi}_{N - 1}^- + s_N^- \varphi_N^-)} \\
                & \times e^{g\bar{\varphi}_N^+ \varphi_{N - 1}^+ - i\delta t V (s_N^+ \bar{\varphi}_N^+ + s^+_{N - 1} \varphi_{N - 1}^+)}\cdots
                  e^{g\bar{\varphi}_1^+ \varphi_0^+ - i \delta t V (s_1^+\bar{\varphi}_1^+ + s_0^+ \varphi_0^+)},
  \end{split}
\end{equation}
where $h = 1-\delta \tau \omega_0 = e^{- \delta \tau \omega_0}$,
$g = 1-i \omega_0 \delta t = e^{- i \omega_0 \delta t}$ and
$\bar{g} = e^{i \omega_0 \delta t}$. Now define the contour time
interval $\dd{t}$ as $\dd{t} = \delta t$ if it is on the 1st (forward)
branch, $\dd{t} = -\delta t$ on the 2nd (backward) branch and
$\dd{t} = -i\delta\tau$ on the 3rd (imaginary) branch. Denote
$\bm{s} = (s_0^+ = s_M^{\sim}, s_1^+, \ldots, s^+_N = s_N^-, \ldots,
s_1^-, s_0^- = s_0^{\sim}, s_1^{\sim}, \ldots, s_{M-1}^{\sim})$, the
influence functional can be written as
\begin{equation}
  I[s(t)] = \frac{1}{Z^{(0)}_E} \int\mathcal{D}[\bar{\bm{\varphi}} \bm{\varphi}]
  e^{-\sum_{j, k = 0}^{2N+M-1} \bar{\varphi}_j S_{j k} \varphi_k - i \dd t V \sum_{k =0}^{2N+M-1} s_k (\varphi_k + \bar{\varphi}_k)},
\end{equation}
where
\begin{equation}
  S = \mqty[
  1 &  &  &  &  &  &  &  &  & - h\\
  - g & 1 &  &  &  &  &  &  &  & \\
  & - g & \ddots &  &  &  &  &  &  & \\
  &  & \ddots & 1 &  &  &  &  &  & \\
  &  &  & - \bar{g} & \ddots &  &  &  &  & \\
  &  &  &  & \ddots & 1 &  &  &  & \\
  &  &  &  &  & - \bar{g} & 1 &  &  & \\
  &  &  &  &  &  & \ddots & \ddots &  & \\
  &  &  &  &  &  &  & - h & 1 & \\
  &  &  &  &  &  &  &  & - h & 1].
\end{equation}
The determinant of $S$ is
\begin{equation}
\det S = 1 + (- 1)^{2 N + M - 1} (- g)^N (- \bar{g})^N (- h)^M = 1-e^{- \beta \omega_0},
\end{equation}
and $[\det S]^{- 1}$ just cancels the free environment partition
function $Z_E^{(0)}$.  As before, the inverse $S^{- 1}$ can be written
in terms of the free environment Green's function. The free
environment Green's function in the Kadanoff contour is defined in a
same form as the Keldysh one \eqref{eq:boson-keldysh-green} that
\begin{equation}
  D_0 (\omega ; t', t'') = -i\expval*{T_{\mathcal{C}}\hat{b}_{\omega} (t')\hat{b}_{\omega}^{\dag} (t'')}_0,
\end{equation}
where the contour $\mathcal{C}$ is referred to the Kadanoff
contour. Then the influence functional has the same form to
\eqref{eq:boson-keldysh-influence-functional-continuous-limit} and the
correlation function is still written as
$\Lambda (t', t'') = iV^2D_0
(\omega_0;t',t'')=i\int\dd{\omega}J(\omega)D_0(\omega;t',t'')$, but
the free environment Green's function $D_0(\omega ; t', t'')$ is now
on the Kadanoff contour.

To be explicit, there are three branches on the Kadanoff contour, and
accordingly the Green's function on the contour can be split into 9
blocks that
\begin{equation}
  D_0 (\omega ; t', t'') = \mqty[
  D_0^{11} (\omega ; t', t'') & D^{12}_0 (\omega ; t', t'') & D_0^{13}(\omega ; t', t'')\\
  D_0^{21} (\omega ; t', t'') & D^{22}_0 (\omega ; t', t'') & D_0^{23}(\omega ; t', t'')\\
  D^{31}_0 (\omega ; t', t'') & D^{32}_0 (\omega ; t', t'') & D^{33}_0(\omega ; t', t'')]
\end{equation}
In the four components $D^{11}_0, D^{12}_0, D^{21}_0, D^{22}_0$, both
arguments $t', t''$ are on the forward or backward branch, thus they
are just the Keldysh Green's functions
\eqref{eq:boson-keldysh-green-pp}-\eqref{eq:boson-keldysh-green-mp}.
Since both arguments of $D^{33}_0$ are on imaginary branch, it is just
the Matsubara Green's function \eqref{eq:boson-imaginary-green} with
an extra factor $i$. The explicit expressions of rest four components,
which have mixed time arguments, are shown as follows.

In $D_0^{13}$ and $D^{23}_0$, the argument $t'$ is on the forward
(backward) branch and $t''$ is on the imaginary branch, which means
that $t''$ always succeeds $t'$ on the contour. Since $t''$ is on the
imaginary branch, we may write it as $t'' = -i\tau''$ and
\begin{equation}
  D_0^{13} (\omega ; t', t'') = D^{23}_0 (\omega ; t', t'') =
  - i\expval*{\hat{b}_{\omega}^{\dag} (-i\tau'') \hat{b}_{\omega} (t')}_0 =
  - in(\omega)e^{\tau'' \omega} e^{- i \omega t'} = -in(\omega) e^{- i \omega (t' - t'')}.
\end{equation}
In $D_0^{31}$ and $D^{32}_0$, the argument $t'$ is on the imaginary branch and
always succeeds $t''$, therefore we may write $t' = -i\tau'$ and
\begin{equation}
  D^{31}_0 (\omega ; t', t'') = D^{32}_0 (\omega ; t', t'') =
  -i\expval*{\hat{b}_{\omega} (-i\tau') \hat{b}^{\dag}_{\omega} (t'')}_0 =
  -i[1 + n(\omega)]e^{- \tau' \omega} e^{i \omega t''} = -i[1 + n(\omega)]e^{- i \omega (t'- t'')} .
\end{equation}

Finally, the influence functional is written in the same form as that
in the Keldysh formalism
\eqref{eq:boson-keldysh-influence-functional-continuous-limit} as
\begin{equation}
  I[s (t)] = e^{- \int_{\mathcal{C}} \dd{t'} \int_{\mathcal{C}} \dd{t''} s(t') \Lambda (t', t'') s(t'')},\quad
  \Lambda (t', t'') = i \int \dd \omega J (\omega) D_0 (\omega ;t',t''),
\end{equation}
and the difference is that $\mathcal{C}$ now stands for the Kadanoff
contour.

\section{System Correlation Functions}\label{sec:boson-system}

In a quantum open system, the information of the system, which can be
characterized by the system correlation functions, is usually of most
interest. In this section, we shall illustrate how to express the system
correlation functions in path integral formalism.

\subsection{On the imaginary-time axis }\label{sec:boson-imaginary-system}

For ease of exposition, let us start from the imaginary-time
formalism again.  At the very first, consider the expectation value of
the system operator $\hat{s}$ that
\begin{equation}
  \expval*{\hat{s}} = \frac{1}{Z} \Tr\relax[e^{- \beta \hat{H}}\hat{s}].
  \label{eq:boson-imaginary-expectation}
\end{equation}
When written in path integral formalism, the expression for
$\Tr\relax[e^{-\beta\hat{H}}\hat{s}]$ is similar to
\eqref{eq:boson-imaginary-path-integral-Z} except that the system
propagator \eqref{eq:boson-imaginary-system-propagator} need to be
modified to
\begin{equation}
  \tilde{K}[\bm{s}] = \mel*{s_M}{e^{- \delta \tau \hat{H}_S}}{s_{M -1}} \cdots
  \mel*{s_1}{e^{- \delta \tau \hat{H}_S} \hat{s}}{s_0},
\end{equation}
where an operator $\hat{s}$ is inserted at the right end. Since
$\hat{s} \ket*{s_0} = s_0 \ket*{s_0}$, we may write
\begin{equation}
  \tilde{K}[\bm{s}] = \mel*{s_M}{e^{- \delta \tau \hat{H}_S}}{s_{M - 1}} \cdots
  \mel*{s_1}{e^{- \delta \tau \hat{H}_S}}{s_0}s_0 = K [\bm{s}] s_0,
\end{equation}
and thus in the continuous limit the path integral expression for
\eqref{eq:boson-imaginary-expectation} is
\begin{equation}
  \expval*{\hat{s}} = \frac{Z^{(0)}_E}{Z} \int\mathcal{D}[s(\tau)] K[s(\tau)]s(0) I[s(\tau)] =
  \frac{1}{Z_S}\int\mathcal{D} [s(\tau)] K[s(\tau)] s(0) I[s(\tau)] .
\end{equation}
From the above expressions, it can be seen that to obtain the
expectation value $\expval*{\hat{s}}$, we need to only insert $s(0)$
in the path integral expression
\eqref{eq:boson-imaginary-path-integral} and then divide it by
$Z_S$. In fact, due to the cyclic property of
\eqref{eq:boson-imaginary-expectation} we can write
$\expval*{\hat{s}} = \expval*{\hat{s} (\tau_1)}$ for any
$0 \leqslant \tau_1 \leqslant \beta$, which means that $\hat{s}$ can
be inserted at any time other than $\tau = 0$ and it yields
\begin{equation}
  \expval*{\hat{s}} = \expval*{\hat{s} (\tau_1)} =
  \frac{1}{Z_S}\int \mathcal{D}[s(\tau)] K[s(\tau)] s(\tau_1) I[s(\tau)], \quad
  0 \leqslant \tau_1 \leqslant \beta .
\end{equation}

The path integral formalism is convenient to express the time-ordered
multiple time correlation functions, since the ordering of the
operators are automatically implemented in the path integral
expressions. For example, let us consider the time-ordered system
correlation function $\expval*{T \hat{s}(\tau_1)
  \hat{s}(\tau_2)}$. Suppose $\tau_1 > \tau_2$, then we have
[$\hat{s}(\tau) = e^{\tau \hat{H}}\hat{s} e^{- \tau \hat{H}}$]
\begin{equation}
  \expval*{T \hat{s}(\tau_1) \hat{s}(\tau_2)} =
  \expval*{\hat{s}(\tau_1) \hat{s} (\tau_2)} =
  \frac{1}{Z} \Tr\relax[e^{-\beta\hat{H}} \hat{s}(\tau_1)\hat{s}(\tau_2)] =
  \frac{1}{Z} \Tr\relax[e^{-(\beta - \tau_1) \hat{H}} \hat{s} e^{-(\tau_1 - \tau_2) \hat{H}} \hat{s}e^{-\tau_2 \hat{H}}].
\end{equation}
To evaluate such a correlation function, we need to insert $\hat{s}$
at time $\tau_1$ and $\tau_2$ in the path integral expression. Suppose
$\tau_1 = j \delta \tau, \tau_2 = k \delta \tau$ with $j > k$, then we
need to modify the system propagator
\eqref{eq:boson-imaginary-system-propagator} to
\begin{equation}
  \tilde{K} [\bm{s}] = \mel*{s_M}{e^{- \delta \tau \hat{H}_S}}{s_{M-1}} \cdots
  \mel*{s_{j + 1}}{e^{- \delta \tau \hat{H}_S} \hat{s}}{s_j} \cdots
  \mel*{s_{k + 1}}{e^{- \delta \tau \hat{H}_S} \hat{s}}{s_k} \cdots
  \mel*{s_1}{e^{- \delta \tau \hat{H}_S}}{s_0} =K[\bm{s}] s_j s_k .
\end{equation}
The corresponding path integral expression is then
\begin{equation}
  \expval*{\hat{s}(\tau_1) \hat{s}(\tau_2)} =
  \frac{1}{Z_S} \int\mathcal{D} [s(\tau)] K[s(\tau)] s(\tau_1) s(\tau_2) I[s(\tau)],
  \quad \tau_1 > \tau_2 .
\end{equation}
When $\tau_1 < \tau_2$, we have
$\expval*{T \hat{s}(\tau_1) \hat{s}(\tau_2)} =
\expval*{\hat{s}(\tau_2) \hat{s}(\tau_1)}$, and following the above
procedure we shall obtain the same path integral expression
\begin{equation}
  \expval*{\hat{s}(\tau_2) \hat{s}(\tau_1)} =
  \frac{1}{Z_S}\int\mathcal{D}[s(\tau)] K[s(\tau)] s(\tau_1) s(\tau_2) I[s(\tau)],
  \quad \tau_1 < \tau_2 .
\end{equation}
Combining the above two expressions, we have the path integral expression for
time-ordered correlation function that
\begin{equation}
  \expval*{T \hat{s}(\tau_1) \hat{s}(\tau_2)} =
  \frac{1}{Z_S}\int\mathcal{D}[s(\tau)] K[s(\tau)] s(\tau_1) s(\tau_2) I[s(\tau)] .
\end{equation}

The evaluation of a non-time-ordered correlation function
$\expval*{\hat{s}(\tau_1) \hat{s}(\tau_2)}$ with $\tau_1 < \tau_2$
needs to consider backward time evolution, thus its formalism needs a
reconstruction. But the basic procedure remains the same and the
details would not be discussed here.

The formalism can be directly generalized to multi-time time-ordered
correlation functions. Basically, we need to only insert all the
corresponding $\hat{s}(\tau)$ at proper times in the path integral
expression \eqref{eq:boson-imaginary-path-integral}. For instance,
\begin{equation}
  \expval*{T \hat{s}(\tau_1) \hat{s}(\tau_2) \hat{s}(\tau_3) \hat{s}(\tau_4)} =
  \frac{1}{Z_S} \int \mathcal{D}[s(\tau)] K[s(\tau)] s(\tau_1) s (\tau_2) s (\tau_3) s (\tau_4) I[s(\tau)] .
\end{equation}

\subsection{On the Keldysh and Kadanoff contour}\label{sec:boson-keldysh-system}

Let us now consider the expectation value $\expval*{\hat{s}(t_1)}$ at
time $t_1$ in the Keldysh formalism, where
$\hat{s} (t_1) = e^{i \hat{H} t_1}\hat{s} e^{- i \hat{H} t_1}$ and
$0 \leqslant t_1 \leqslant t_f$. This quantity can be evaluated as
\begin{equation}
  \expval*{\hat{s} (t_1)} = \frac{1}{Z (t_f)}
  \Tr\relax[e^{- i \hat{H}(t_f - t_1)} \hat{s} e^{- i \hat{H} t_1} \hat{\rho} (0) e^{i \hat{H} t_f}],
\end{equation}
or
\begin{equation}
  \expval*{\hat{s} (t_1)} = \frac{1}{Z (t_f)}
  \Tr\relax[e^{- i \hat{H}t_f} \hat{\rho} (0) e^{i \hat{H} t_1} \hat{s} e^{i \hat{H} (t_f - t_1)}].
\end{equation}
If the former expression is adopted, the operator $\hat{s}$ is
inserted into the forward branch at time step $t_1$, then we have
\begin{equation}
  \expval*{\hat{s}(t_1)} = \frac{1}{Z_S} \int\mathcal{D}[s(t)] K[s(t)] s(t_1^+) I[s(t)] .
\end{equation}
Equivalently, if the latter expression is adopted, then the operator
$\hat{s}$ is inserted into the backward branch that
\begin{equation}
  \expval*{\hat{s}(t_1)} = \frac{1}{Z_S} \int\mathcal{D}[s(t)] K[s(t)] s(t_1^-) I[s(t)] .
\end{equation}

In section \ref{sec:boson-imaginary-system}, it is mentioned that in
imaginary time formalism the backward evolution needs to be considered
for non-time-ordered correlation function and thus a reconstruction of
the formalism is needed. In Keldysh formalism the backward evolution
has been already taken into consideration, so the formalism need not
to be revised for two-time correlation functions. Take the quantity
$\expval*{\hat{s} (t_1) \hat{s} (t_2)}$ with
$0 \leqslant t_1, t_2 \leqslant t_f$ as an example. If $t_1 > t_2$, we
can evaluate this quantity in two ways that
\begin{equation}
  \expval*{\hat{s} (t_1) \hat{s} (t_2)} =
  \frac{1}{Z} \Tr\relax[\hat{s}(t_1) \hat{s} (t_2) \hat{\rho} (0)] =
  \frac{1}{Z} \Tr\relax[e^{- i \hat{H}(t_f - t_1)} \hat{s} e^{- i \hat{H} (t_1 - t_2)} \hat{s} e^{- i \hat{H} t_2}
  \hat{\rho} (0) e^{i \hat{H} t_f}],
\end{equation}
and
\begin{equation}
  \expval*{\hat{s} (t_1) \hat{s} (t_2)} =
  \frac{1}{Z} \Tr\relax[\hat{s}(t_2) \hat{\rho} (0) \hat{s} (t_1)] =
  \frac{1}{Z} \Tr\relax[e^{- i \hat{H}(t_f - t_2)} \hat{s} e^{- i \hat{H} t_2} \hat{\rho} (0) e^{i \hat{H} t_1}
  \hat{s} e^{i \hat{H} (t_f - t_1)}] .
\end{equation}
The corresponding path integral expressions are
\begin{equation}
  \expval*{\hat{s} (t_1) \hat{s} (t_2)} = \frac{1}{Z_S} \int\mathcal{D}[s(t)] K[s(t)] s(t_1^+) s(t_2^+) I[s(t)],
\end{equation}
and
\begin{equation}
  \expval*{\hat{s} (t_1) \hat{s} (t_2)} = \frac{1}{Z_S} \int\mathcal{D}[s(t)] K[s(t)] s(t_1^-) s(t_2^+) I[s(t)] .
  \label{eq:boson-keldysh-t1t2-1}
\end{equation}
If $t_1 < t_2$, this quantity can be also evaluated in two ways that
\begin{equation}
  \expval*{\hat{s} (t_1) \hat{s} (t_2)} =
  \frac{1}{Z} \Tr\relax [\hat{s}(t_2) \hat{\rho}(0) \hat{s}(t_1)] =
  \frac{1}{Z} \Tr\relax [e^{- i \hat{H}(t_f - t_2)} \hat{s} e^{- i \hat{H} t_2} \hat{\rho} (0) e^{i \hat{H} t_1}
  \hat{s} e^{i \hat{H} (t_f - t_1)}],
\end{equation}
and
\begin{equation}
  \expval*{\hat{s} (t_1) \hat{s} (t_2)} =
  \frac{1}{Z} \Tr\relax[\hat{\rho} (0) \hat{s} (t_1) \hat{s} (t_2)] =
  \frac{1}{Z} \Tr\relax[e^{- i \hat{H} t_f} \hat{\rho} (0) e^{i \hat{H} t_1} \hat{s} e^{i \hat{H}(t_2 - t_1)}
  \hat{s} {e^{i \hat{H} (t_f - t_2)}}].
\end{equation}
The corresponding path integral expressions are
\begin{equation}
  \expval*{\hat{s} (t_1) \hat{s} (t_2)} = \frac{1}{Z_S}\int\mathcal{D}[s(t)] K[s(t)] s(t_1^-) s(t_2^+) I[s(t)],
  \label{eq:boson-keldysh-t1t2-2}
\end{equation}
and
\begin{equation}
  \expval*{\hat{s} (t_1) \hat{s} (t_2)} = \frac{1}{Z_S}\int\mathcal{D}[s (t)] K[s(t)] s(t_1^-) s(t_2^-) I[s(t)] .
\end{equation}
It can be seen that \eqref{eq:boson-keldysh-t1t2-1} in $t_1 > t_2$
case and \eqref{eq:boson-keldysh-t1t2-2} in $t_1 < t_2$ case are the
same, which indicates that it is mostly convenient to insert
$\hat{s}(t_1)$ on the backward branch and $\hat{s}(t_2)$ on the
forward branch. By doing so, the chronology of $t_1$ and $t_2$ becomes
irrelevant and can be taken out of consideration.

It is mostly convenient to express the correlation function in
contour-ordered form, since the operators are automatically ordered in
the path integral formalism on the contour. The quantities
$\expval*{\hat{s} (t_1) \hat{s} (t_2)}$ and
$\expval*{\hat{s} (t_2) \hat{s} (t_1)}$ are in fact components
$\expval*{T_{\mathcal{C}} \hat{s}(t_1^-) \hat{s}(t_2^+)}$ and
$\expval*{T_{\mathcal{C}} \hat{s}(t_1^+) \hat{s}(t_2^-)}$ of
$\expval*{T_{\mathcal{C}} \hat{s}(t_1) \hat{s}(t_2)}$,
respectively. To evaluate
$\expval*{T_{\mathcal{C}} \hat{s} (t_1) \hat{s} (t_2)}$, we just
insert the operators into the proper position on the contour and
obtain
\begin{equation}
  \expval*{T_{\mathcal{C}}\hat{s}(t_1) \hat{s}(t_2)} = \frac{1}{Z_S} \int\mathcal{D}[s(t)] K[s(t)] s(t_1) s(t_2) I[s(t)].
\end{equation}

For Kadanoff formalism, the expression for contour-ordered two-time
correlation function has the same form as above. The difference
between Keldysh and Kadanoff formalism is that the Kadanoff contour
has an extra imaginary branch, therefore we can obtain the
imaginary-time correlation function by simply inserting two operators
into the imaginary contour. We can also obtain the mixed-time
correlation function by inserting one operator into the real-time
branch and another to the imaginary-time branch.

The expression can be generalized to multi-time contour-ordered
correlation function directly. For instance, the four-time
contour-ordered correlation function
$\expval*{T_{\mathcal{C}} \hat{s}(t_1) \hat{s}(t_2) \hat{s}(t_3)\hat{s}(t_4)}$ in
both Keldysh and Kadanoff formalism can be written as
\begin{equation}
  \expval*{T_{\mathcal{C}} \hat{s}(t_1) \hat{s}(t_2) \hat{s}(t_3) \hat{s}(t_4)}=
  \frac{1}{Z_S} \int\mathcal{D}[s(t)] K[s(t)] s(t_1) s(t_2) s(t_3) s(t_4) I[s(t)].
\end{equation}

\section{Generating Functional Method}
\label{sec:boson-generating-functional}

In the path integral formalism, although the environment degrees of
freedom is integrated out, we can still retrieve the environment
information from the system correlation function via the generating
functional method
\cite{popovic2021-quantum,gribben2021-using,chen2023-heat}. In this
section, we shall illustrate how this is done.

\subsection{On the imaginary-time axis}\label{sec:boson-imaginary-system-environment}

First, let us consider the expectation value of the system environment
coupling energy $\expval*{\hat{H}_{S E}}$ in thermal equilibrium. The
quantity
$\expval*{\hat{H}_{S E}} = \frac{1}{Z} \Tr\relax[e^{- \beta
  \hat{H}}\hat{H}_{S E}]$ can be written as
\begin{equation}
  \expval*{\hat{H}_{S E}} = V \expval*{\hat{s} \hat{b}} +
  V\expval*{\hat{s} \hat{b}^{\dag}} = 2 V \Re \expval*{\hat{s}\hat{b}} .
\end{equation}
Due to the cyclic property, we can also write it as
\begin{equation}
  \expval*{\hat{H}_{S E}} = \expval*{\hat{H}_{S E} (\tau_1)} =
  2 V\Re \expval*{\hat{s} (\tau_1) \hat{b} (\tau_1)}, \quad
  0 \leqslant \tau_1 \leqslant \beta .
\end{equation}
The operator $\hat{s}$ can be inserted at any position on the imaginary-time
axis and yields an extra term $s (\tau_1)$ in the path integral expression.
The operator $\hat{b}$ acts on the environment, thus it can not be inserted
into the system propagator, and we need to use generating functional method to
handle it.

To employ the generating functional method, we insert a source term
$e^{-\xi V \hat{b}}$ in \eqref{eq:boson-imaginary-partition} at time
$\tau_1 = k \delta \tau$ and modify the partition function to
\begin{equation}
  \begin{split}
    Z^{\xi}  = & \sum_{s_0, \ldots, s_{N - 1}} \int \mathcal{D}[\bar{\varphi}_0\varphi_0] e^{-\bar{\varphi}_0 \varphi_0} \cdots
                 \int \mathcal{D}[\bar{\varphi}_{M-1}\varphi_{M-1}] e^{-\bar{\varphi}_{M-1} \varphi_{M-1}}\\
               & \times \mel*{s_M\varphi_M}{e^{- \delta \tau \hat{H}}}{s_{M-1}\varphi_{M-1}} \cdots
                 \mel*{s_{k + 1}\varphi_{k+1}}{e^{- \delta \tau \hat{H}}e^{- \xi V \hat{b}}\hat{s}}{s_k\varphi_k} \cdots
                 \mel*{s_1 \varphi_1}{e^{-\delta \tau \hat{H}}}{s_0\varphi_0},
  \end{split}
\end{equation}
where the boundary condition is
$\bra*{s_M\varphi_M} = \bra*{s_0\varphi_0}$. Then
$V \expval*{\hat{s}(\tau_1) \hat{b}(\tau_1)}$ can be evaluated as
\begin{equation}
  V \expval*{\hat{s} (\tau_1) \hat{b} (\tau_1)} = - \frac{1}{Z} \eval{
  \frac{\delta Z^{\xi}}{\delta \xi}}_{\xi = 0} .
\end{equation}
With this source term, we have
\begin{equation}
  \mel*{s_{k + 1} \varphi_1}{e^{- \delta \tau \hat{H}} e^{- \xi V \hat{b}}\hat{s}}{s_k \varphi_k} =
  \mel*{s_{k+1}}{e^{- \delta \tau \hat{H}_S}}{s_k}
  e^{g\bar{\varphi}_{k+1}\varphi_k - \delta\tau V[s_{k+1} \bar{\varphi}_{k+1} + (s_k + \delta\tau^{-1} \xi) \varphi_k]}.
\end{equation}
The quantity $\delta \tau^{- 1}$ behaves as a Dirac delta function at
time $\tau_1$, and the final path integral expression writes
\begin{equation}
  Z^{\xi} = Z^{(0)}_E \int \mathcal{D}[s(\tau)] K[s(\tau)] s(\tau_1) I^{\xi}[s(\tau)],
\end{equation}
where
\begin{equation}
  \begin{split}
    I^{\xi} [s (\tau)] = & e^{-\int_0^{\beta}\dd{\tau'}\int_0^{\beta}\dd{\tau''}[s(\tau')+
                           \xi\delta(\tau'-\tau_1)]\Lambda(\tau',\tau'')s(\tau'')}\\
    = & e^{-\int_0^{\beta}\dd{\tau'}\int_0^{\beta}\dd{\tau''} s(\tau') \Lambda (\tau', \tau'') s (\tau'') -
        \xi \int_0^{\beta} \dd{\tau'}\Lambda (\tau_1, \tau') s (\tau')} .
  \end{split}
\end{equation}
Differentiation of $Z^{\xi}$ with respect to $\xi$ yields
\begin{equation}
  \begin{split}
    V \expval*{\hat{s}(\tau_1) \hat{b} (\tau_1)}
    = & \frac{1}{Z_S} \int\mathcal{D}[s(\tau)] \qty{K[s(\tau)] s(\tau_1) I[s(\tau)] \times
       \int_0^{\beta} \dd{\tau'} \Lambda(\tau_1, \tau') s(\tau')}\\
    = & \int_0^{\beta} \dd{\tau'} \qty{ \frac{1}{Z_S} \int\mathcal{D}[s(\tau)]
        K[s(\tau)] I[s(\tau)] \Lambda (\tau_1, \tau') s (\tau') s(\tau_1)} \\
      = & \int_0^{\beta} \dd{\tau'} \Lambda (\tau_1, \tau') \expval*{T \hat{s}(\tau') \hat{s}(\tau_1)}.
  \end{split}
\end{equation}
For continuous bosons case, following the same procedure we have
\begin{equation}
  \sum_k V_k \expval*{\hat{s} (\tau_1) \hat{b}_k (\tau_1)}=
  \int_0^{\beta}\dd{\tau'} \Lambda (\tau_1, \tau') \expval*{T \hat{s}(\tau') \hat{s} (\tau_1)}.
\end{equation}

Now let us consider the path integral expression of the environment
time-ordered Green's function
\begin{equation}
  D(\omega_0 ; \tau_1, \tau_2) = -\expval*{T \hat{b} (\tau_1) \hat{b}^{\dag}(\tau_2)}.
\end{equation}
Here we suppose $\tau_1 > \tau_2$ ($\tau_1 < \tau_2$ case follows the
same procedure and yields the same result for time-ordered Green's
function), then we have
\begin{equation}
  -\expval*{T \hat{b} (\tau_1) \hat{b}^{\dag} (\tau_2)} =
  -\expval*{\hat{b} (\tau_1) \hat{b}^{\dag} (\tau_2)} =
  -\frac{1}{Z} \Tr\relax[e^{-\beta\hat{H}} \hat{b}(\tau_1) \hat{b}^{\dag}(\tau_2)] .
\end{equation}
This expression can be also evaluated using the generating functional
method.  Suppose $\tau_1 = j \delta \tau, \tau_2 = k \delta \tau$ with
$j > k$, we can insert two source terms
$e^{-\xi_1 \hat{b}},e^{-\xi_2 \hat{b}^{\dag}}$ at $\tau_1,\tau_2$ in
\eqref{eq:boson-imaginary-partition} and modify it as
\begin{equation}
  \begin{split}
    Z^{\xi_1 \xi_2} = & \sum_{s_0, \ldots, s_{N-1}}\int \mathcal{D}[\bar{\varphi}_0\varphi_0] e^{- \bar{\varphi}_0 \varphi_0} \cdots
                        \int \mathcal{D} [\bar{\varphi}_{M-1} \varphi_{M-1}] e^{-\bar{\varphi}_{M-1} \varphi_{M-1}}\\
                      & \times \mel*{s_M\varphi_M}{e^{- \delta \tau \hat{H}}}{s_{M-1}\varphi_{M-1}} \cdots
                        \mel*{s_{j + 1} \varphi_{j + 1}}{e^{-\delta \tau \hat{H}} e^{- \xi_1 \hat{b}}}{s_j \varphi_j} \\
                      & \times\cdots\mel*{s_{k + 1}\varphi_{k + 1}}{e^{-\xi_2\hat{b}^{\dag}}e^{- \delta \tau \hat{H}}}{s_k\varphi_k}
                       \cdots \mel*{s_1 \varphi_1}{e^{- \delta \tau \hat{H}}}{s_0\varphi_0},
  \end{split}
\end{equation}
then the correlation function can be evaluated as
\begin{equation}
  -\expval*{\hat{b}(\tau_1) \hat{b}^{\dag}(\tau_2)} =
  -\frac{1}{Z}\eval{\frac{\delta^2 Z^{\xi_1 \xi_2}}{\delta \xi_1 \delta \xi_2}}_{\xi_1 = \xi_2 = 0}.
\end{equation}
With the source terms, we have
\begin{equation}
  \mel*{s_{j+1} \varphi_{j+1}}{e^{-\delta\tau\hat{H}} e^{- \xi_1\hat{b}}}{s_j \varphi_j} =
  \mel*{s_{j+1}}{e^{-\delta\tau\hat{H}_S}}{s_j}
  e^{g \bar{\varphi}_{j+1} \varphi_j - \delta\tau V [s_{j+1} \bar{\varphi}_{j+1} + (s_j + \delta \tau^{- 1} V^{-1} \xi_1) \varphi_j]},
\end{equation}
and
\begin{equation}
  \mel*{s_{k+1}\varphi_{k+1}}{e^{-\xi_2\hat{b}^{\dag}}e^{-\delta\tau\hat{H}}}{s_k\varphi_k} =
  \mel*{s_{k + 1}}{e^{- \delta \tau\hat{H}_S}}{s_k}
  e^{g \bar{\varphi}_{k+1}\varphi_k - \delta\tau V [(s_{k+1} + \delta \tau^{-1} V^{-1} \xi_2) \bar{\varphi}_{k+1} + s_k\varphi_k]}.
\end{equation}
The quantity $\delta \tau^{- 1}$ in the above two expressions can be
treated as two Dirac delta functions at time $\tau_1$ and $\tau_2$
respectively, and the final path integral expression for
$Z^{\xi_1 \xi_2}$ is then
\begin{equation}
  Z^{\xi_1 \xi_2} = Z^{(0)}_E \int \mathcal{D}[s(\tau)] K[s(\tau)] I^{\xi_1\xi_2}[s(\tau)],
\end{equation}
where
\begin{equation}
  \begin{split}
    I^{\xi_1\xi_2}[s(\tau)]
    = & e^{-\int_0^{\beta}\dd{\tau'}\int_0^{\beta}\dd{\tau''}
        [s (\tau') + V^{- 1}\xi_1\delta(\tau' - \tau_1)]
        \Lambda (\tau', \tau'') [s(\tau'') + V^{- 1} \xi_2 \delta(\tau''- \tau_2)]} \\
    = & e^{-\int_0^{\beta}\dd{\tau'}\int_0^{\beta} \dd{\tau''} s(\tau') \Lambda (\tau', \tau'') s(\tau'') 
        -\xi_1V^{-1} \int_0^{\beta}\dd{\tau'} \Lambda(\tau_1, \tau') s(\tau')
        - \xi_2V^{-1}\int_0^{\beta} \dd{\tau'} s(\tau') \Lambda(\tau', \tau_2)
        - \xi_1\xi_2V^{- 2} \Lambda (\tau_1, \tau_2)}.
  \end{split}
\end{equation}
Differentiation of $Z^{\xi_1 \xi_2}$ with respect to $\xi_1, \xi_2$
yields
\begin{equation}
  \begin{split}
    D(\omega_0 ; \tau_1, \tau_2)
    = & V^{- 2} \qty[ \Lambda (\tau_1, \tau_2)
        -\int_0^{\beta} \dd{\tau'}\int_0^{\beta}\dd{\tau''} \expval*{T \hat{s}(\tau') \hat{s}(\tau'')}
        \Lambda (\tau_1, \tau') \Lambda (\tau'',\tau_2)]\\
    = & D_0(\omega_0 ; \tau_1, \tau_2) - V^2\int_0^{\beta}\dd{\tau'}\int_0^{\beta}\dd{\tau''}
        D_0 (\omega ; \tau_1, \tau') \expval*{T \hat{s}(\tau') \hat{s}(\tau'')} D_0(\omega_0 ; \tau'', \tau_2),
  \end{split}
  \label{eq:boson-imaginary-environment-green-function-single}
\end{equation}
where $D_0$ is the free environment Green's function. It can be seen
from the above expression that the term
$- V^2\expval*{T \hat{s} (\tau') \hat{s}(\tau'')}$ acts as the
$T$-matrix of the environment Green's function.

For continuous bosons case, we are usually interested in the Green's function
\begin{equation}
  D(\omega_k ; \tau_1, \tau_2) = - \expval*{T \hat{b}_{\omega_k}(\tau_1) \hat{b}_{\omega_k}(\tau_2)},
\end{equation}
whose expression is similar to
\eqref{eq:boson-imaginary-environment-green-function-single} that
\begin{equation}
  D(\omega_k ; \tau_1, \tau_2) = D_0(\omega_k ; \tau_1, \tau_2)
  -V_k^2\int_0^{\beta}\dd{\tau'}\int_0^{\beta}\dd{\tau''}D_0 (\omega_k ;\tau_1, \tau')
  \expval*{T s(\tau') s(\tau'')} D_0 (\omega_k ;\tau'', \tau_2).
\end{equation}

\subsection{On the Keldysh and Kadanoff contour}

In Keldysh and Kadanoff formalism, the quantity
$\expval*{\hat{H}_{S E} (t_1)}$ can be evaluated as
\begin{equation}
  \expval*{\hat{H}_{S E} (t_1)} = V [\expval*{\hat{s} (t_1) \hat{b}(t_1)}
    + \expval*{\hat{s} (t_1) \hat{b}^{\dag} (t_1)}] = 2\Re V\expval*{\hat{s} (t_1) \hat{b} (t_1)}.
 \end{equation}
 To evaluate the quantity $V \expval*{\hat{s} (t_1) \hat{b} (t_1)}$,
 we insert $\hat{s}$ and a source term $e^{\xi V \hat{b}}$ at $t_1$ on
 the contour. For instance, we may insert them at $t_1^+ = k \delta t$
 that
\begin{equation}
  \mel*{s_{k + 1}^+ \varphi^+_{k + 1}}{e^{- i \hat{H} \dd t} \hat{s}e^{\xi V\hat{b}}}{s_k^+ \varphi_k^+} =
  \mel*{s_{k + 1}^+}{e^{- i \hat{H}_S\dd{t}}}{s_k^+}s_k^+
  e^{g \bar{\varphi}_{k+1}^+ \varphi^+_k - i V\dd{t}[s_{k+1}^+ \bar{\varphi}_{k+1}^+ + (s_k^+ +i\xi\dd{t}^{-1}) \varphi_k^+]}.
\end{equation}
Here $\dd{t} = \delta t$ on the forward branch, and when it is on the
backward or imaginary branch we should have $\dd{t} = - \delta t$ or
$\dd{t} = - i \delta \tau$. The quantity $\dd{t}^{- 1}$ can be treated
as the Dirac delta function on the contour. Then the quantity
$V \expval*{\hat{s}(t_1) \hat{b}(t_1)}$ can be evaluated as
\begin{equation}
  V\expval*{\hat{s}(t_1) \hat{b}(t_1)}=\frac{1}{Z}\eval{\frac{\delta Z^{\xi}}{\delta \xi}}_{\xi = 0} .
\end{equation}

After integrating out the environment degrees of freedom, we should have
\begin{equation}
  Z^{\xi}(t_f) = Z^{(0)}_E\int\mathcal{D} [s(t)] K[s(t)] s(t_1) I^{\xi}[s(t)],
\end{equation}
where
\begin{equation}
  \begin{split}
    I^{\xi} [s (t)]
    = & e^{-\int_{\mathcal{C}}\dd{t'} \int_{\mathcal{C}}\dd{t''} [s(t') + i\xi\delta_{\mathcal{C}}(t' - t_1)] \Lambda(t', t'') s(t'')}\\
    = & e^{-\int_{\mathcal{C}}\dd{t'} \int_{\mathcal{C}}\dd{t''} s(t') \Lambda(t', t'') s(t'') - i\xi\int_{\mathcal{C}} \dd{t'} \Lambda (t_1, t') s(t')}.
  \end{split}
\end{equation}
Here $\delta_{\mathcal{C}}(t' - t_1)$ is the contour Dirac delta
function, which becomes the usual Dirac delta function with an extra
sign, when $t'$ coincides with $t_1$. The sign of
$\delta_{\mathcal{C}}$ is $+1$ when $t_1$ is on the forward branch,
$-1$ on the backward branch and $i$ on the imaginary branch. It should
be noted that here the sign of $\delta_{\mathcal{C}}(t' - t_1)$ would
cancel the sign of $\dd{t'}$.

Then differentiation of $Z^{\xi}$ with respect to $\xi$ yields
\begin{equation}
  \begin{split}
    V \expval*{\hat{s} (t_1) \hat{b} (t_1)}
    = & -\frac{i}{Z_S} \int\mathcal{D} [s (t)]\qty{K[s(t)] s(t_1) I[s(t)]\times \int\dd{t'} \Lambda (t_1, t') s (t')}\\
    = & -i\int_{\mathcal{C}} \dd{t'} \Lambda(t_1, t') \expval*{T_{\mathcal{C}}\hat{s} (t')\hat{s}(t_1)}.
  \end{split}
  \label{eq:boson-keldysh-system-environement-coupling}
\end{equation}
In the continuous bosons case, we just replace the left-handed side of
the above expression by
$\sum_k V_k \expval*{\hat{s} (t_1) \hat{b}_k (t_1)}$ that
\begin{equation}
  \sum_k V_k \expval*{\hat{s} (t_1) \hat{b}_k (t_1)} =
  -i\int_{\mathcal{C}} \dd{t'} \Lambda (t_1, t') \expval*{T_{\mathcal{C}}\hat{s}(t') \hat{s} (t_1)}.
\end{equation}

For multiple baths case in Keldysh formalism, we should have
\begin{equation}
  V_{\alpha} \expval*{\hat{s} (t_1) \hat{b}_{\alpha} (t_1)} =
  -i \int_{\mathcal{C}}\dd{t'}\Lambda_{\alpha} (t_1, t')
  \expval*{T_C \hat{s} (t') \hat{s} (t_1)}, \quad
  \Lambda_{\alpha}(t_1, t') = i\int\dd{\omega} J_{\alpha}(\omega) D_{\alpha}(\omega ; t_1, t').
\end{equation}
where $D_{\alpha}$ is the free Green's function for $\alpha$th bath. In
continuous bosons case, it becomes
\begin{equation}
  \sum_k V_{\alpha k} \expval*{\hat{s} (t_1) \hat{b}_{\alpha k} (t_1)} =
  -i\int_{\mathcal{C}} \dd{t'} \Lambda_{\alpha}(t_1, t')\expval*{T_C \hat{s} (t')\hat{s} (t_1)}. 
\end{equation}

The environment contour-ordered Green's function is defined as
\begin{equation}
  D(\omega_0 ; t_1, t_2) = -i\expval*{T_{\mathcal{C}} \hat{b} (t_1) \hat{b}^{\dag} (t_2)}.
\end{equation}
To evaluate it, we insert two sources $e^{\xi_1\hat{b}}$ and
$e^{\xi_2 \hat{b}^{\dag}}$ at $t_1$ and $t_2$ on the contour. After
integrating out the environment, we should obtain that
\begin{equation}
  Z^{\xi_1 \xi_2}(t_f) = Z^{(0)}_E \int\mathcal{D}[s(t)] K[s(t)] I^{\xi_1 \xi_2}[s(t)],
\end{equation}
where
\begin{equation}
  \begin{split}
    I^{\xi_1 \xi_2} [s (t)]
    = & e^{-\int_{\mathcal{C}}\dd{t'}\int_{\mathcal{C}}\dd{t''} [s(t') + i V^{- 1}\xi_1\delta_{\mathcal{C}} (t' - t_1)]
        \Lambda (t', t'') [s (t'') + iV^{- 1} \xi_2 \delta_{\mathcal{C}}(t'' - t_2)]} \\
    = & e^{-\int_{\mathcal{C}}\dd{t'}\int_{\mathcal{C}}\dd{t''}s (t') \Lambda (t', t'') s(t'')
        -i\xi_1V^{- 1}\int_{\mathcal{C}} \dd{t'} \Lambda(t_1, t') s (t')
        -i\xi_2V^{- 1}\int_{\mathcal{C}} \dd{t'} s(t')\Lambda (t', t_2) + \xi_1\xi_2V^{- 2}\Lambda (t_1, t_2)}.
  \end{split}
\end{equation}
Here the signs of $\delta_{\mathcal{C}}(t' - t_1)$ and
$\delta_{\mathcal{C}}(t'' - t_2)$ cancel the signs of $\dd{t'}$ and
$\dd{t''}$, respectively. The environment Green's function is then
evaluated as
\begin{equation}
  D(\omega_0 ; t_1, t_2) = -\frac{i}{Z} \eval{\frac{\delta^2 Z^{\xi_1\xi_2}}{\delta \xi_1 \delta \xi_2}}_{\xi_1 = \xi_2 = 0},
\end{equation}
which is
\begin{equation}
  \begin{split}
    D(\omega_0 ; t_1, t_2)
    = & - iV^{-2}\Lambda (t_1, t_2) + i V^{- 2}\int_{\mathcal{C}} \dd{t'} \int_{\mathcal{C}} \dd {t''}
        \Lambda (t_1, t') \expval*{T_C\hat{s}(t') \hat{s}(t'')} \Lambda(t'', t_2)\\
    = & D_0(\omega_0 ; t_1, t_2) - iV^2\int_{\mathcal{C}}\dd{t'}\int_{\mathcal{C}}\dd{t''}
        D_0(\omega_0 ; t_1, t') \expval*{T_C \hat{s}(t') \hat{s}(t'')} D_0(\omega_0 ; t'', t_2).
  \end{split}
\end{equation}
Here the term $-iV^2 \expval*{T_C \hat{s} (t') \hat{s} (t'')}$ acts as
the $T$-matrix for environment Green's function. In continuous bosons
case, we have
\begin{equation}
  D(\omega_k ; t_1, t_2)=D_0 (\omega_k ; t_1, t_2) - iV_k^2\int_{\mathcal{C}}\dd{t'} \int_{\mathcal{C}}\dd{t''}
  D_0(\omega_k ; t_1, t') \expval*{T_C \hat{s} (t')\hat{s} (t'')}D_0 (\omega_k ; t'', t_2).
\end{equation}

\subsection{Heat currents}

In nonequilibrium dynamics, sometimes the heat current between the
system and environment is of interest
\cite{popovic2021-quantum,chen2023-heat}. This nonequilibrium dynamics
can be conveniently expressed in the Keldysh formalism. The heat
current that flows out from the environment is defined as
\begin{equation}
  \mathcal{I}(t_1) = -\expval*{\hat{\dot{H}}_E (t_1)} =
  -i\expval*{[\hat{H}, \hat{H}_E (t_1)]} =
  -iV\omega_0 \expval*{\hat{s}(t_1) [\hat{b}(t_1) - \hat{b}^{\dag}(t_1)]} .
\end{equation}
Since $\expval*{\hat{s}(t_1) \hat{b}(t_1)}$ and
$\expval*{\hat{s} (t_1)\hat{b}^{\dag}(t_1)}$ are conjugate to each
other, we can write
\begin{equation}
  \mathcal{I} (t_1) = 2\Im V\omega_0\expval*{\hat{s} (t_1) \hat{b}(t_1)}.
\end{equation}
According to \eqref{eq:boson-keldysh-system-environement-coupling}, we
have immediately
\begin{equation}
  \begin{split}
    \mathcal{I}(t_1) &= -2\Re\qty[\int_{\mathcal{C}}\dd{t'}V^2\omega_0 D_0(\omega_0;t_1,t')\expval*{T_{\mathcal{C}}\hat{s}(t') \hat{s}(t_1)}]\\
    & =-2\Re\qty[\int_{\mathcal{C}} \dd{t'}\int\dd{\omega} J(\omega) \omega D_0(\omega ; t_1, t')\expval*{T_C \hat{s}(t') \hat{s}(t_1)}].
  \end{split}
  \label{eq:boson-keldysh-current}
\end{equation}
The heat currents expression for multiple baths case has the same form
to \eqref{eq:boson-keldysh-current} that the current
$\mathcal{I}_{\alpha}$ flowing out from $\alpha$th is
\begin{equation}
  \begin{split}
  \mathcal{I}_{\alpha}(t_1) &= -2\Re\qty[\int_{\mathcal{C}}\dd{t'}V_{\alpha}^2 \omega_{\alpha}D_{\alpha}(\omega_{\alpha} ; t_1, t')
                              \expval*{T_{\mathcal{C}}\hat{s} (t') \hat{s} (t_1)}]\\
    &=-2 \Re \qty[ \int_{\mathcal{C}} \dd{t'}
  \int\dd{\omega} J_{\alpha} (\omega) \omega D_{\alpha}(\omega ; t_1, t')
      \expval*{T_{\mathcal{C}}\hat{s} (t') \hat{s} (t_1)}].
  \end{split}
\end{equation}

\section{Fermion Path Integral Formalism}\label{sec:fermion-path-integral}

In the fermion case, we shall consider the so-called Toulouse model
\cite{leggett1987-dynamics} or Anderson-Fano model
\cite{mahan2000-many}. The system is a quantum dot with Hamiltonian
$\hat{H}_S = \varepsilon_d \hat{a}^{\dag} \hat{a}$, and the
environment consists of free fermions that hybridize between the
system and environment for which
\begin{equation}
  \hat{H}_E = \sum_k \varepsilon_k \hat{c}_k^{\dag} \hat{c}_k, \quad
  \hat{H}_{S E} = \sum_k V_k (\hat{a}^{\dag} \hat{c}_k + \hat{c}_k^{\dag}\hat{a}).
\end{equation}
The environment is characterized by the spectral function
\begin{equation}
  \Gamma(\varepsilon) = \sum_k V_k^2 \delta(\varepsilon - \varepsilon_k).
  \label{eq:fermion-spectral-continuous}
\end{equation}

When two flavors of spin and the Coulomb interaction are taken into
consideration, this model is generalized to the famous Anderson
impurity model \cite{anderson1961-localized,mahan2000-many}. If
multiple orbitals are considered, it is generalized further to the
Slater-Kanamori model
\cite{kanamori1963-electron,georges2013-strong}. The influence
functional remains the same form for these models.

As in the boson case, we shall consider the single fermion situation
first, where
\begin{equation}
  \hat{H}_E = \varepsilon_0\hat{c}^{\dag}\hat{c}, \quad \hat{H}_{S E} = V(\hat{a}^{\dag} \hat{c} + \hat{c}^{\dag} \hat{a}).
\end{equation}
This single fermion case corresponds to the spectral function
\begin{equation}
  \Gamma(\varepsilon)=V^2\delta(\varepsilon-\varepsilon_0).
  \label{eq:fermion-spectral-single}
\end{equation}
To restore the continuous fermions case, we need to only replace the
spectral function \eqref{eq:fermion-spectral-single} by
\eqref{eq:fermion-spectral-continuous}.

\subsection{On the imaginary-time axis}

With the aid of the fermion coherent state, the basic idea of deriving
the fermion path integral formalism is the same as that of boson
formalism shown in section \ref{sec:boson-path-integral}. Here we use
the imaginary-time path integral formalism to illustrate the essential
ingredients.

In thermal equilibrium, the whole density matrix is written as
$\hat{\rho}=e^{- \beta \hat{H}}$. Spliting $\beta = M \delta \tau$
with $M \rightarrow \infty$ yields the partition function as
\begin{equation}
  Z = \Tr\relax[e^{- \delta \tau \hat{H}} \cdots e^{-\delta \tau \hat{H}}].
\end{equation}
Inserting the fermion coherent identity operator
\eqref{eq:fermion-coherent-identity} between every two exponents and
employing the fermion coherent trace
\eqref{eq:fermion-coherent-trace}, we shall have
\begin{equation}
  Z = \int \mathcal{D} [\bar{\bm{a}}\bm{a}] e^{-\bar{\bm{a}} \bm{a}} \int \mathcal{D}[\bar{\bm{c}}\bm{c}] e^{-\bar{\bm{c}}\bm{c}}
  \mel*{(- a_M) (- c_M)}{e^{- \delta \tau \hat{H}}}{a_{M - 1}c_{M - 1}} \cdots \mel*{a_1c_1}{e^{- \delta \tau \hat{H}}}{a_0c_0}
\end{equation}
with the boundary condition $\bra*{(- a_M) (- c_M)} = \bra*{ (- a_0) (-c_0)}$, where
\begin{equation}
  \int \mathcal{D}[\bar{\bm{a}} \bm{a}] e^{-\bar{\bm{a}} \bm{a}} =
  \int \dd{\bar{a}_0} \dd{a_0} e^{- \bar{a}_0 a_0} \cdots \int \dd{\bar{a}_{M - 1}} \dd{a_{M - 1}} e^{- \bar{a}_{M-1} a_{M-1}},
\end{equation}
and
\begin{equation}
  \int \mathcal{D}[\bar{\bm{c}}\bm{c}]e^{-\bar{\bm{c}} \bm{c}} =
  \int \dd{\bar{c}_0} \dd{c_0}e^{- \bar{c}_0 c_0} \cdots \int\dd{\bar{c}_{M-1}} \dd{c_{M-1}}e^{-\bar{c}_{M-1} c_{M-1}} .
\end{equation}
Employing the Trotter-Suzuki decomposition, we have
\begin{equation}
  \mel*{a_{k+1} c_{k+1}}{e^{- \delta \tau \hat{H}}}{a_k c_k} =
  e^{ (1-\delta \tau \varepsilon_d)\bar{a}_{k+1} a_k} e^{g\bar{c}_{k+1} c_k - \delta \tau V (\bar{a}_{k+1} c_k +\bar{c}_{k+1} a_k)},
\end{equation}
where
$g = 1 - \delta \tau \varepsilon_0 = e^{- \delta \tau \varepsilon_0}$.
Accordingly, with boundary condition
$\bar{c}_M = \bar{c}_0, \bar{a}_M = \bar{a}_0$ we may write the
partition function in path integral formalism as
\begin{equation}
  Z = Z^{(0)}_E \int\mathcal{D}[\bar{\bm{a}}\bm{a}] K[\bar{\bm{a}}\bm{a}] I[\bar{\bm{a}}\bm{a}],
\end{equation}
where the bare system propagator is
\begin{equation}
  K[\bar{\bm{a}} \bm{a}] = e^{- \sum_k \bar{a}_k a_k}e^{-(1 - \delta\tau\varepsilon_d)\bar{a}_M a_{M-1} }
  e^{ (1 - \delta \tau \varepsilon_d)\bar{a}_{M-1} a_{M-2}} \cdots e^{(1 - \delta\tau \varepsilon_d)\bar{a}_1 a_0}.
  \label{eq:femion-system-propagator}
\end{equation}
Since
\begin{equation}
  -\bar{a}_k a_k +  (1 - \delta \tau\varepsilon_d)\bar{a}_k a_{k-1} =
  -\delta\tau\bar{a}_k [\delta \tau^{-1} (a_k - a_{k-1}) + \varepsilon_d a_{k-1}],
\end{equation}
and in the continuous limit $\delta \tau \rightarrow 0$ we may denote
$(a_k-a_{k-1}) / \delta\tau$ as $\partial_{\tau'} a (\tau')$, then the
bare system propagator can be written symbolically as
\begin{equation}
  K[\bar{a}(\tau) a(\tau)] = e^{-\int_0^{\beta}\dd{\tau'}\bar{a} (\tau')\qty(\frac{\partial}{\partial \tau'} + \varepsilon_d) a(\tau')}
\end{equation}
with boundary condition $\bar{a}(\tau) = -\bar{a}(0)$. It should be
noted that it is sometimes more convenient to define the system
propagator \eqref{eq:femion-system-propagator} without the term
$e^{-\sum_k\bar{a}_ka_k}$, see
Ref. \cite{chen2024-gtempo,xu2024-grassmann} for example. In that
case, this term is absorbed in the definition of the measure
$\mathcal{D}[\bar{\bm{a}}\bm{a}]$.

The influence functional writes
\begin{equation}
  I[\bar{\bm{a}}\bm{a}]  =  \frac{1}{Z^{(0)}_E} \int\mathcal{D} [\bar{\bm{c}} \bm{c}]e^{-\bar{\bm{c}} \bm{c}}
  e^{- g \bar{c}_M c_{M - 1} +\delta \tau V (\bar{a}_M c_{M - 1} + \bar{c}_M a_{M - 1})} \cdots
  e^{g \bar{c}_1 c_0 - \delta \tau V (\bar{a}_1 c_0 + \bar{c}_1a_0)},
\end{equation}
or we can write it in the Gaussian integral form as
\begin{equation}
  I[\bar{\bm{a}}\bm{a}] = \frac{1}{Z^{(0)}_E} \int\mathcal{D}[\bar{\bm{c}}\bm{c}]
  e^{-\sum_{j, k = 0}^{M-1}\bar{c}_j S_{j k} c_k - \delta \tau V \sum_{k = 0}^{M - 1}(\bar{a}_{k + 1} c_k + \bar{c}_{k + 1} a_k)}, 
  \label{eq:boson-influence-functional-before-integration}
\end{equation}
where the matrix $S$ is
\begin{equation}
  S = \mqty[
    1 &  &  &  & h\\
    - h & 1 &  &  & \\
    & - h & \ddots &  & \\
    &  &  & 1 & \\
    &  &  & - h & 1].
\end{equation}

Note that in
$-\delta\tau V \sum_{k = 0}^{M - 1} (\bar{a}_{k + 1} c_k+\bar{c}_{k +
  1} a_k)$ of
\eqref{eq:boson-influence-functional-before-integration}, there is one
step displacement in the subscript of $\bar{a}_{k + 1} c_k$ and
$\bar{c}_{k + 1} a_k$. In the continuous limit
$\delta \tau \rightarrow 0$, the error caused by this displacement is
of higher order, thus for convenience we can evaluate the influence
functional as
\begin{equation}
  I[\bar{\bm{a}} \bm{a}] = \frac{1}{Z^{(0)}_E} \int
  \mathcal{D} [\bar{\bm{c}} \bm{c}] e^{- \sum_{j, k = 0}^{M -
  1} \bar{c}_j S_{j k} c_k - \delta \tau V \sum_{k = 0}^{M - 1}
  (\bar{a}_k c_k + \bar{c}_k a_k)} .
\end{equation}
The determinant of $S$ is simply
$\det S = 1 + (-1)^{M-1} (-h)^{M-1}h = 1 + h^M = 1 +
e^{-\beta\varepsilon_0}$, which cancels $Z^{(0)}_E$. The inverse of
$S$ is
\begin{equation}
  S^{- 1} = [1 - f (\varepsilon_0)] \mqty[
    1 & - h^{M - 1} & \cdots & - h^2 & - h\\
    h & 1 &  & - h^3 & - h^2\\
    h^2 & h & \ddots & \vdots & \vdots\\
    \vdots & \vdots &  & 1 & - h^{M - 1}\\
    h^{M - 1} & h^{M - 2} & \cdots & h & 1],
\end{equation}
where $f (\varepsilon) = (e^{\beta \varepsilon} + 1)^{-1}$ is the
Fermi-Dirac distribution function. The inverse matrix $S^{-1}$, as in
the boson case, can be expressed by the Matsubara Green's
function. The free environment Matsubara Green's function in the
fermion case is defined as
\begin{equation}
  G_0(\varepsilon ; \tau', \tau'') = -\expval*{T \hat{c}_{\varepsilon}(\tau') \hat{c}_{\varepsilon}^{\dag}(\tau'')}_0 =
  \begin{cases}
    -\expval*{\hat{c}_{\varepsilon} (\tau') \hat{c}_{\varepsilon}^{\dag}(\tau'')}_0, & \tau' \geqslant \tau'' ;\\
    \expval*{\hat{c}_{\varepsilon}^{\dag} (\tau'') \hat{c}_{\varepsilon}(\tau')}_0, & \tau' < \tau'',
  \end{cases}
  \label{eq:fermion-imaginary-green}
\end{equation}
where $\hat{c}_{\varepsilon} (\tau) = \hat{c} e^{- \tau \varepsilon},
\hat{c}^{\dag}_{\varepsilon} (\tau) = \hat{c}^{\dag} e^{\tau \varepsilon}$ in
free environment. Note that in the definition of fermion Green's function, a
minus sign arises when two arguments interchange their chronology. The
explicit form of $G_0$ is
\begin{equation}
  G_0 (\varepsilon ; \tau', \tau'') = \begin{cases}
    - [1 - f (\varepsilon)] e^{- (\tau' - \tau'') \varepsilon}, & \tau'\geqslant \tau'';\\
    f (\varepsilon) e^{- (\tau' - \tau'') \varepsilon}, & \tau' < \tau''.
  \end{cases}
\end{equation}
Therefore in the continuous limit $\delta \tau \rightarrow 0$, we have
\begin{equation}
  Z_S = \int \mathcal{D}[\bar{a}(\tau) a(\tau)] K[\bar{a}(\tau) a(\tau)] I[\bar{a}(\tau) a(\tau)],
\end{equation}
where
\begin{equation}
  I[\bar{a}(\tau) a(\tau)] = e^{-\int_0^{\beta}\dd{\tau'}\int_0^{\beta}\dd{\tau''}\bar{a}(\tau') \Delta(\tau', \tau'') a(\tau'')} .
\end{equation}
Here $\Delta (\tau', \tau'')$ is usually called the hybridization
function that
\begin{equation}
  \Delta(\tau', \tau'') = V^2 G_0(\varepsilon_0 ; \tau', \tau'')=
  \int\dd{\varepsilon}\Gamma(\varepsilon) G_0(\varepsilon ; \tau', \tau'').
\end{equation}

\subsection{On the Keldysh contour}

Now let us consider the fermion path integral expression in Keldysh
formalism. As in the boson case, suppose at initial time $t = 0$, the
whole system is in a product state that the density matrix is
$\hat{\rho} (0) = \hat{\rho}_S (0) \otimes \hat{\rho}_E$. The
environment is assumed in thermal equilibrium that
$\hat{\rho}_E = e^{- \beta \hat{H}_E}$. The density matrix at time
$t_f$ is given by
$\hat{\rho} (t_f) = e^{- i \hat{H} t_f} \hat{\rho}(0) e^{i
  \hat{H}t_f}$.  Splitting $t_f = N \delta t$ with
$N \rightarrow \infty$ gives the whole partition function as
\begin{equation}
  Z(t_f) = \Tr\relax[e^{- i \hat{H} \delta t} \cdots e^{- i \hat{H} \delta t} \hat{\rho} (0)
  e^{i \hat{H} \delta t} \cdots e^{i \hat{H} \delta t}].
\end{equation}
Inserting the fermion coherent identity operator
\eqref{eq:fermion-coherent-identity} between every two exponents and
employing the coherent trace \eqref{eq:fermion-coherent-trace}, we can
write it in the path integral formalism as
\begin{equation}
  Z (t_f) = Z^{(0)}_E \int \mathcal{D}[\bar{\bm{a}} \bm{a}] K[\bar{\bm{a}}\bm{a}] I[\bar{\bm{a}}\bm{a}],
\end{equation}
where the system propagator is (with the boundary condition
$\bra*{-a_N^+}=\bra*{-a_N^-}$)
\begin{equation}
  K[\bar{\bm{a}}\bm{a}] = e^{- \sum_k \bar{a}^{\pm}_ka_k^{\pm}}
  \mel*{-a_N^+}{e^{- i \hat{H}_S \delta t}}{a_{N - 1}^+} \cdots
  \mel*{a_1^+}{e^{- i \hat{H}_S \delta t}}{a_0^+}
  \mel*{a_0^+}{\hat{\rho}_S (0)}{a_0^-}
  \mel*{a_0^-}{e^{i \hat{H}_S\delta t}}{a_1^-} \cdots
  \mel*{a_{N - 1}^-}{e^{i \hat{H}_S \delta t}}{a_N^-}.
\end{equation}
On the forward branch, we have
\begin{equation}
  e^{- \bar{a}_k^+ a_k^+}  \mel*{a_k^+}{e^{- i \hat{H}_S \delta t}}{a_{k - 1}^+} =
  e^{- \delta t \bar{a}_k^+ [\delta t^{- 1} (a_k^+- a_{k - 1}^+) + i \varepsilon_d a^+_{k - 1}]},
\end{equation}
and on the backward branch, we have
\begin{equation}
  e^{-\bar{a}_k^- a_k^-} \mel*{a_k^-}{e^{i \hat{H}_S \delta t}}{a_{k+1}^-} =
  e^{-\delta t \bar{a}_k [\delta t^{-1}(a_k^- -a_{k+1}^-) - i\varepsilon_d a_{k+1}^-]} =
  e^{\delta t\bar{a}_k [\delta t^{-1} (a_{k+1}^- - a_k^-) + i\varepsilon_d a_{k+1}^-]}. 
\end{equation}
Recall that $\dd{t} = \delta t$ on the forward branch and
$\dd{t} = - \delta t$ on the backward branch, and define the contour
derivative as
$\frac{\partial a(t)}{\partial t} = \frac{1}{\dd{t}} [a(t + \dd{t}) -
a(t)]$, where $a(t + \dd{t})$ means the variable an infinitesimal step
ahead of $a(t)$ on the contour. To be explicit, the contour
derivatives on the forward and backward branches are respectively
\begin{equation}
  \frac{\partial a (t^+)}{\partial t} = \frac{1}{\delta t} (a_k^+ - a^+_{k -1}), \quad
  \frac{\partial a (t^-)}{\partial t} = \frac{1}{(-\delta t)}(a^-_k - a^-_{k + 1}) =
  \frac{1}{\delta t}(a^-_{k + 1} - a^-_k) .
  \label{eq:keldysh-contour-derivative}
\end{equation}
Then in the continuous limit, we may write $K$ symbolically in terms
of contour derivative as
\begin{equation}
  K[\bar{a} (t) a (t)] = e^{- \int_C \dd t' a (t') \qty(\frac{\partial}{\partial t'} + i \varepsilon_d)a (t')}
  \rho_S(\bar{a}_0^+, a_0^-), \quad
  \rho_S(\bar{a}^+_0, a_0^-) = \mel*{a_0^+}{\hat{\rho}_S (0)}{a_0^-}.
\end{equation}

The influence functional writes
\begin{equation}
  \begin{split}
    I[\bar{\bm{a}} \bm{a}]
    = & \frac{1}{Z^{(0)}_E} \int\mathcal{D} [\bar{\bm{c}} \bm{c}]e^{-\bar{\bm{c}} \bm{c}}
        e^{- g \bar{c}_N^+c_{N-1}^+ + i \delta t V (\bar{a}_N^+ c^+_{N-1} + \bar{c}^+_Na_{N - 1}^+)} \cdots
        e^{g \bar{c}_1^+ c^+_0 - i \delta t V(\bar{a}_1^+ c_0^+ + \bar{c}_1^+ a_0^+)} \\
      & \times \mel*{c_0^+}{\hat{\rho}_E}{c_0^-}
        e^{\bar{g}\bar{c}^-_0 c_1^- + i \delta t V (\bar{a}_0^- c_1^- +\bar{c}_0^- a_1^-)} \cdots
        e^{\bar{g} \bar{c}^-_{N-1} c_N^-+ i \delta t V (\bar{a}_{N-1}^- c_N^- + \bar{c}_{N-1}^-a_N^-)},
\end{split}
\end{equation}
where $g = 1 - i \varepsilon_0 \delta t = e^{- i \varepsilon_0 \delta t}$
and $\bar{g}$ is its complex conjugate. Here we first evaluate the
term $\mel*{c^+_0}{\hat{\rho}_E}{c_0^-}$. Split the inverse
temperature in $\hat{\rho}_E$ into $M \rightarrow \infty$ pieces that
$\beta = M \delta \tau$, we have
\begin{equation}
  \begin{split}
    \mel*{c_0^+}{\hat{\rho}_E}{c_0^-}
    = & \int\mathcal{D}[\bar{c}^{\sim}_1 c_1^{\sim}] \cdots
        \int\mathcal{D} [\bar{c}_{M-1}^{\sim} c_{M - 1}^{\sim}]e^{-\sum_{k=1}^{M-1}\bar{c}_k^{\sim} c^{\sim}_k}
        \mel*{c_0^+}{e^{- \delta \tau\hat{H}_E}}{c_{M - 1}^{\sim}} \cdots
        \mel*{c_1^{\sim}}{e^{- \delta\tau \hat{H}_E}}{c_0^+}\\
    = & \int\mathcal{D}[\bar{c}^{\sim}_1 c_1^{\sim}] \cdots
        \int\mathcal{D}[\bar{c}_{M - 1}^{\sim} c_{M - 1}^{\sim}]
        e^{-\sum_{k = 1}^{M - 1} \bar{c}_k^{\sim} c^{\sim}_k} e^{h \bar{c}_0^+c^{\sim}_{M-1}}
        e^{h \bar{c}^{\sim}_{M - 1} c^{\sim}_{M - 2}} \cdots e^{h \bar{c}^{\sim}_1 c_0^+},
  \end{split}
\end{equation}
where $h = 1 - \delta \tau \varepsilon_0 = e^{- \delta \tau \varepsilon_0}$.
Then we can write
\begin{equation}
  \mel*{c_0^+}{\hat{\rho}_E}{c_0^-} =
  \int\mathcal{D}[\bar{c}^{\sim}_1c_1^{\sim}] \cdots
  \int\mathcal{D}[\bar{c}_{M-1}^{\sim} c_{M-1}^{\sim}]
  e^{- \sum_{j k} \bar{c}_j^{\sim} S_{j k}c^{\sim}_k + h \bar{c}_0^+ c^{\sim}_{M - 1} + h \bar{c}_1^{\sim}c_0^+},
\end{equation}
where $S$ and its inverse are
\begin{equation}
  S = \mqty[
  1 &  &  &  &  & \\
  - h & 1 &  &  &  & \\
  & - h & \ddots &  &  & \\
  &  &  & \ddots &  & \\
  &  &  & - h & 1 & \\
  &  &  &  & - h & 1], \quad
  S^{- 1} = \mqty[
  1 &  &  &  &  & \\
  h & 1 &  &  &  & \\
  h^2 & h & \ddots &  &  & \\
  \vdots & \vdots & \ddots & \ddots &  & \\
  h^{M - 3} & h^{M - 4} & \cdots & h & 1 & \\
  h^{M - 2} & h^{M - 3} & \cdots & \cdots & h & 1].
\end{equation}
Thus we have
\begin{equation}
  \mel*{c_0^+}{\hat{\rho}_E}{c_0^-} = e^{h^2 \bar{c}_0^+ S^{-1}_{M - 1, 1} c_0^-} = e^{e^{- \beta \varepsilon_0} \bar{c}_0^+ c_0^-}.
\end{equation}

Now denote $\bm{a} = (a_0^+, a_1^+, \ldots, a^+_N = - a_N^-, \ldots,a_1^-, a_0^-)$
and neglect the displacement in the subscript, the influence functional can be written as
\begin{equation}
  I[\bar{\bm{a}} \bm{a}] = \frac{1}{Z_E^{(0)}}\int\mathcal{D} [\bar{\bm{c}} \bm{c}]
  e^{-\sum_{j,k = 0}^{2N}\bar{c}_j S_{j k} c_k - i \dd{t} V \sum_{k = 0}^{2N} (\bar{a}_kc_k + \bar{c}_k a_k)},
\end{equation}
where the matrix $S$ is
\begin{equation}
  S = \mqty[
    1 &  &  &  &  &  &  &  & - e^{- \beta \varepsilon_0}\\
    - g & 1 &  &  &  &  &  &  & \\
    & - g & \ddots &  &  &  &  &  & \\
    &  &  & 1 &  &  &  &  & \\
    &  &  & g & 1 &  &  &  & \\
    &  &  &  & - \bar{g} & 1 &  &  & \\
    &  &  &  &  &  & \ddots &  & \\
    &  &  &  &  &  &  & 1 & \\
    &  &  &  &  &  &  & - \bar{g} & 1].
\end{equation}
The determinant of $S$ is
\begin{equation}
  \det S = 1 + (- 1)^{2 N} (- g)^{N - 1} g (- \bar{g})^N (- e^{- \beta\varepsilon_0}) = 1 + e^{- \beta \varepsilon_0},
\end{equation}
which is just the free environment partition function $Z^{(0)}_E$. The
inverse $S^{-1}$ is
[$f = f (\varepsilon_0) = (e^{\beta \varepsilon_0} + 1)^{- 1}$]
\begin{equation}
  \mqty[
    1 - f & - f \bar{g} & \cdots & - f \bar{g}^{N - 1} & f\bar{g}^N & f \bar{g}^{N - 1} & \cdots & f \bar{g} & f\\
    (1 - f) g & 1 - f &  & \vdots & f \bar{g}^{N - 1} & f \bar{g}^{N-2} &  & f & f g\\
    \vdots & (1 - f) g & \ddots & - f \bar{g} & \vdots & \vdots &  &\vdots & \vdots\\
    (1 - f) g^{N - 1} &  &  & 1 - f & f \bar{g} & f &  & f g^{N - 2} & fg^{N - 1}\\
    - (1 - f) g^N & - (1 - f) g^{N - 1} &  & - (1 - f) g & 1 - f & - fg &  &- fg^{N - 1} & - f g^N\\
    - (1 - f) g^{N - 1} & \vdots &  & - (1 - f) & (1 - f) \bar{g} & 1 - f &  & \vdots & - f g^{N - 1}\\
    \vdots & - (1 - f) g &  & - (1 - f) \bar{g} & (1 - f)\bar{g}^2 & (1 - f) \bar{g} &  & - f g & \vdots\\
    - (1 - f) g & - (1 - f) &  & \vdots & \vdots &  & \ddots & 1 - f & - f g\\
    - (1 - f) & - (1 - f) \bar{g} & \cdots & - (1 - f) \bar{g}^{N-1} & (1-f) \bar{g}^N &
    (1 - f) \bar{g}^{N - 1} & \cdots & (1 -f) \bar{g} & (1 - f)].
\end{equation}

Now we try to express $S^{-1}$ in terms of Keldysh Green's
function. The free environment contour-ordered Green's function is
defined as
\begin{equation}
  G_0(\varepsilon ; t', t') = -i\expval*{T_{\mathcal{C}} \hat{c}_{\varepsilon} (t')\hat{c}_{\varepsilon}^{\dag} (t'')}_0 .
\end{equation}
To be explicit, we have
\begin{equation}
  G^{11}_0(\varepsilon ; t', t'') = - i\expval*{T \hat{c}_{\varepsilon} (t')\hat{c}_{\varepsilon}^{\dag}  (t'')}_0 =
  \begin{cases}
    -i[1 - f (\varepsilon)] e^{- i \varepsilon (t' - t'')}, & t' > t'' ;\\
    i f(\varepsilon) e^{- i \varepsilon (t' - t'')}, & t' < t'',
  \end{cases}
  \label{eq:fermion-keldysh-green-pp}
\end{equation}
\begin{equation}
  G^{12}_0 (\varepsilon ; t', t'') = i \expval*{\hat{c}_{\varepsilon}^{\dag}(t'') \hat{c}_{\varepsilon} (t')}_0 =
  i f (\varepsilon) e^{- i\varepsilon (t' - t'')},
  \label{eq:fermion-keldysh-green-pm}
\end{equation}
\begin{equation}
  G^{21}_0 (\varepsilon ; t', t'') = - i \expval*{\hat{c}_{\varepsilon} (t')\hat{c}_{\varepsilon}^{\dag} (t'')}_0 =
  -i [1 - f (\varepsilon)]e^{- i \varepsilon (t' - t'')},
  \label{eq:fermion-keldysh-green-mp}
\end{equation}
\begin{equation}
  G^{22}_0 (\varepsilon ; t', t'') = - i \expval*{\bar{T}\hat{c}_{\varepsilon} (t') \hat{c}_{\varepsilon}^{\dag} (t'')}_0 =
  \begin{cases}
    i f (\varepsilon) e^{- i \varepsilon (t' - t'')}, & t' > t'';\\
    - i [1 - f (\varepsilon)]e^{- i \varepsilon (t' - t'')}, & t' < t'',
  \end{cases}
  \label{eq:fermion-keldysh-green-mm}
\end{equation}
where $f (\varepsilon) = (e^{\beta \varepsilon} + 1)^{- 1}$. Finally,
we have
\begin{equation}
  S^{- 1} = i \mqty[
    G_0^{11} & - G_0^{12}\\
    - G_0^{21} & G_0^{22}],
\end{equation}
and in the continuous limit $\delta t \rightarrow 0$ the influence functional
can be written as
\begin{equation}
  I[\bar{a}(t) a (t)] = e^{-\int_{\mathcal{C}} \dd{t'} \int_{\mathcal{C}} \dd{t''}\bar{a}(t') \Delta (t', t'') a (t'')},
  \label{eq:fermion-keldysh-influence-functional}
\end{equation}
where the hybridization function is
\begin{equation}
  \Delta (t', t'') = i\mathcal{P}_{t' t''} V^2 G_0 (\varepsilon_0 ; t', t'') =
  i\mathcal{P}_{t' t''} \int \dd{\varepsilon} \Gamma(\varepsilon) G_0(\varepsilon ; t', t'').
\end{equation}
Here $\mathcal{P}_{t' t''} = 1$ when $t', t''$ on the same contour,
otherwise $\mathcal{P}_{t' t''} = - 1$. It should be noted the
existence of $\mathcal{P}_{t' t''}$ means that sign issue can not be
handled solely by the definition of the fermion Green's function.

If the environment consists of multiple baths, then the hybridization
function becomes
\begin{equation}
  \Delta(t', t'') = \sum_{\alpha}\Delta_{\alpha}(t', t''),
\end{equation}
where $\Delta_{\alpha}$ is the hybridization function due to $\alpha$th bath
for which
\begin{equation}
  \Delta_{\alpha}(t', t'') = i\mathcal{P}_{t' t''} V^2_{\alpha} G_{\alpha}(\varepsilon_{\alpha} ; t', t'') =
   i\mathcal{P}_{t' t''} \int \dd{\varepsilon}\Gamma_{\alpha}(\varepsilon) G_{\alpha}(\varepsilon; t', t''). 
\end{equation}
Here $\Gamma_{\alpha}$ and $G_{\alpha}$ are the spectral function and
free Green's function for the the $\alpha$th bath.

\subsection{On the Kadanoff contour}

Suppose the whole system is initially in thermal equilibrium that the whole
density matrix is $\hat{\rho} (0) = e^{- \beta \hat{H}}$, then at time $t_f$
the partition function is
\begin{equation}
  Z (\beta, t_f) = \Tr\relax [e^{- i \hat{H} t_f} e^{- \beta \hat{H}} e^{i\hat{H} t_f}] =
  \Tr\relax [e^{- \beta \hat{H}} e^{i \hat{H} t_f} e^{-i\hat{H} t_f}].
\end{equation}
Splitting $t_f = N \delta t, \beta = M \delta \tau$ with
$N \rightarrow \infty, M \rightarrow \infty$, then the partition
function writes
\begin{equation}
  \begin{split}
    Z(\beta, t_f) =
    & Z^{(0)}_E \int \mathcal{D}[\bar{\bm{a}}\bm{a}] e^{- \bar{\bm{a} \bm{}} \bm{a}}
      \int\mathcal{D} [\bar{\bm{c}} \bm{c}] e^{-\bar{\bm{c}} \bm{c}}
      \mel*{(- a^{\sim}_M) (- c^{\sim}_M)}{e^{- \delta \tau \hat{H}}}{a^{\sim}_{M - 1} c^{\sim}_{M - 1}}\cdots
      \mel*{a^{\sim}_1 c^{\sim}_1}{e^{- \delta \tau \hat{H}}}{a^{\sim}_0c^{\sim}_0}\\
    & \times \mel*{a_0^- c_0^-}{e^{i \hat{H} \delta t}}{a_1^- c_1^-} \cdots
      \mel*{a_{N - 1}^- c_{N - 1}^-}{e^{i \hat{H} \delta t}}{a_N^- c_N^-} 
      \mel*{a_N^+ c_N^+}{e^{- i \hat{H} \delta t}}{a^+_{N - 1}c_{N - 1}^+} \cdots
      \mel*{a_1^+ c_1^+}{e^{- i \hat{H} \delta t}}{a_0^+ c_0^+},
  \end{split}
\end{equation}
where the boundary condition is
$\bra*{(-a^{\sim}_M) (-c^{\sim}_M)} = \bra*{(- a_0^+) (- c_0^+)},
\bra*{a_0^- c_0^-} = \bra*{a^{\sim}_0 c^{\sim}_0}, \bra*{a_N^+ c_N^+}
= \bra*{a_N^- c_N^- }$. It can be written in the path integral form as
\begin{equation}
Z(t_f) = Z^{(0)}_E \int \mathcal{D} [\bar{\bm{a}}\bm{a}] K[\bar{\bm{a}} \bm{a}] I[\bar{\bm{a}}\bm{a}],
\end{equation}
where the system propagator writes
\begin{equation}
  \begin{split}
    K [\bar{\bm{a}} \bm{a}]  =
    & \mel*{(- a^{\sim}_M)}{e^{-\delta \tau \hat{H}_S}}{a^{\sim}_{M-1}} \cdots
      \mel*{a^{\sim}_1}{e^{- \delta \tau \hat{H}_S}}{a^{\sim}_0}
      \mel*{a_0^-}{e^{i\hat{H}_S \delta t}}{a_1^-}\\
    & \times \cdots \expval*{a_{N - 1}^-}{e^{i \hat{H}_S \delta t}}{a_N^-}
      \mel*{a_N^+}{e^{- i \hat{H}_S \delta t}}{a_{N - 1}^+}\cdots
      \mel*{a_1^+}{e^{- i \hat{H}_S \delta t}}{a_0^+}.
  \end{split}
\end{equation}
Define $\dd{t} = \delta t$ on the forward branch, $\dd{t} = - \delta t$ on
the backward branch and $\dd{t} = - i \delta \tau$ on the imaginary branch.
Accordingly, the contour derivative on the imaginary branch is
\begin{equation}
  \frac{\partial a (t^{\sim})}{\partial t} = \frac{1}{-i\delta\tau}(a_k^{\sim} - a^{\sim}_{k-1}),
\end{equation}
and the contour derivative on the other two branches is the same as
\eqref{eq:keldysh-contour-derivative}. Then in the continuous limit,
we may write the propagator symbolically as
\begin{equation}
  K [\bar{a}(t) a(t)] = e^{- \int_C \dd t' \bar{a} (t') \qty(\frac{\partial}{\partial t} + i \varepsilon_d) a (t')}.
\end{equation}

The influence functional writes
\begin{equation}
  \begin{split}
    I[\bar{\bm{a}} \bm{a}] =
    & \frac{1}{Z^{(0)}_E} \int\mathcal{D} [\bar{\bm{c}} \bm{c}]e^{-\bar{\bm{c}} \bm{c}}
      e^{-h\bar{c}^{\sim}_M c^{\sim}_{M-1} + \delta\tau V(\bar{c}^{\sim}_M a^{\sim}_{M-1} + \bar{a}^{\sim}_M c^{\sim}_{M-1})} \cdots
      e^{h \bar{c}^{\sim}_1 c^{\sim}_0 - \delta \tau V(\bar{c}_1^{\sim} a^{\sim}_0 + \bar{a}^{\sim}_1 c^{\sim}_0)}\\
    & \times e^{\bar{g} \bar{c}_0^- c_1^- + i \delta t V(\bar{c}_0^- a_1^- + \bar{a}_0^- c_1^-)} \cdots
      e^{\bar{g}\bar{c}_{N - 1}^- c_N^- + i \delta t V (\bar{c}_{N - 1}^- a_N^- +\bar{a}_{N - 1}^- c_N^-)} \\
    & \times e^{g \bar{c}_N^+ c_{N - 1}^+ - i \delta t V(\bar{c}_N^+ a_{N - 1}^+ + \bar{a}_N^+ c_{N - 1}^+)} \cdots
      e^{g\bar{c}_1^+ c_0^+ - i \delta t V (\bar{c}_1^+ a_0^+ +\bar{a}_1^+ c_0^+)}.
  \end{split}
\end{equation}
Denote $\bm{c} = (c_0^+ = c^{\sim}_M, c_1^+, \ldots, c^+_N = c_N^-,
\ldots, c_1^-, c_0^- = c^{\sim}_0, c^{\sim}_1, \ldots, c^{\sim}_{M - 1})$ and
neglect the displacement in the subscripts, the influence functional can be
written in the form as
\begin{equation}
 I[\bar{\bm{c}} \bm{c}] = e^{-\sum_{j, k = 0}^{2N+M-1} \bar{c}_j S_{j k} c_k - i \dd{t} V \sum_{k = 0}^{2N+M-1}(\bar{c}_k a_k + \bar{a}_k c_k)},
\end{equation}
where the matrix $S$ is
\begin{equation}
  S = \mqty[
  1 &  &  &  &  &  &  &  &  & h\\
  - g & 1 &  &  &  &  &  &  &  & \\
  & - g & \ddots &  &  &  &  &  &  & \\
  &  &  & 1 &  &  &  &  &  & \\
  &  &  & - \bar{g} & \ddots &  &  &  &  & \\
  &  &  &  &  & 1 &  &  &  & \\
  &  &  &  &  & - \bar{g} & 1 &  &  & \\
  &  &  &  &  &  &  & \ddots &  & \\
  &  &  &  &  &  &  & - h & 1 & \\
  &  &  &  &  &  &  &  & - h & 1].
\end{equation}
The determinant of $S$ is
\begin{equation}
  \det S = 1 + (- 1)^{2 N + M - 1} (- g)^N (- \bar{g})^N (- h)^{M - 1} h = 1 + e^{- \beta \varepsilon_0},
\end{equation}
which is just the free environment partition function $Z^{(0)}_E$. The inverse
of $S$, as before, can be expressed by the contour-ordered Green's function.
The free environment contour-ordered Green's function is defined as
\begin{equation}
  G_0 (\varepsilon ; t', t'') = -i\expval*{T_{\mathcal{C}}\hat{c}_{\varepsilon} (t')\hat{c}^{\dag}_{\varepsilon} (t'')}_0,
\end{equation}
where $T_{\mathcal{C}}$ is the Kadanoff contour order operator. There
are three branches on the contour, and thus $G_0$ can be split into 9
blocks that
\begin{equation}
  G_0 (\varepsilon ; t', t'') = \mqty[
    G^{11}_0 (\varepsilon ; t', t'') & G^{12}_0 (\varepsilon ; t', t'') & G^{13}_0 (\varepsilon ; t', t'')\\
    G^{21} (\varepsilon ; t', t'') & G^{22}_0 (\varepsilon ; t', t'') & G^{23}_0 (\varepsilon ; t', t'')\\
    G^{31}_0 (\varepsilon ; t', t'') & G^{32}_0 (\varepsilon ; t', t'') & G^{33}_0 (\varepsilon ; t', t'')].
\end{equation}

In $G^{11}_0, G^{12}_0, G^{21}_0$ and $G^{22}$, both arguments
$t', t''$ are on real-time branches, and they are just the Keldysh
Green's functions
\eqref{eq:fermion-keldysh-green-pp}-\eqref{eq:fermion-keldysh-green-mm}. The
arguments $t', t''$ of $G^{33}_0$ are on imaginary-time branch, and
$G^{33}_0$ is just the Matsubara Green's function
\eqref{eq:fermion-imaginary-green} with an extra factor $i$. The
explicit form of the rest four components of $G_0$ are listed as
follows:
\begin{equation}
  G^{13}_0 (\varepsilon ; t', t'') = G_0^{23} (\varepsilon ; t', t'') =
  -i[1- f (\varepsilon)] e^{- i \varepsilon t'} e^{\tau'' \varepsilon_0} =
  -i[1- f (\varepsilon)] e^{- i \varepsilon (t' - t'')},
\end{equation}
\begin{equation}
  G^{31}_0 (\varepsilon ; t', t'') = G^{32} (\varepsilon ; t', t'') =
  if(\varepsilon) e^{- \tau' \varepsilon_0} e^{i \varepsilon_0 t''} =
  if(\varepsilon) e^{- i(t' - t'')}.
\end{equation}
Finally, the influence functional can be written in the same form in
the Keldysh contour \eqref{eq:fermion-keldysh-influence-functional} as
\begin{equation}
  I[\bar{a}(t) a(t)] = e^{-\int_{\mathcal{C}} \dd{t'} \int_{\mathcal{C}} \dd{t''}\bar{a} (t') \Delta (t', t'') a (t'')},
\end{equation}
where the hybridization function is
\begin{equation}
  \Delta(t', t'') = iV^2G_0(\varepsilon_0 ; t', t'') =
  i \int \dd \varepsilon \Gamma (\varepsilon) G_0(\varepsilon ; t', t'').
\end{equation}

Note that unlike that in the Keldysh formalism, here the hybridization
function $\Delta (t', t'')$ does not contain an extra sign factor
$\mathcal{P}_{t', t''}$. This indicates that the sign issue in the
Kadanoff formalism is again handled by the definition of the
contour-ordered Green's function.

\subsection{System correlation functions}

The operator of a physical observable involves pairs of annihilation
and creation operators, e.g., the system particle number operator is
$\hat{a}^{\dag} \hat{a}$. On the imaginary-time axis, the corresponding
expectation value is
$\expval*{\hat{a}^{\dag} (\tau_1) \hat{a} (\tau_1)}$ with
$0 \leqslant \tau_1 \leqslant \beta$, which can be expressed as the
time-ordered correlation function as
\begin{equation}
  \expval*{\hat{a}^{\dag} (\tau_1) \hat{a} (\tau_1)} = - \expval*{T \hat{a} (\tau_1) \hat{a}^{\dag} (\tau_1 + 0)}.
\end{equation}
The time-ordered correlation function
$\expval*{T \hat{a} (\tau_1)\hat{a}^{\dag} (\tau_2)}$ can be
conveniently expressed by the path integral formalism, where the time
ordering is automatically implemented. As in the boson case, we need
to just insert $\hat{a}$ at $\tau_1$ and $\hat{a}^{\dag}$ at $\tau_2$,
which yields
\begin{equation}
  \expval*{T \hat{a} (\tau_1) \hat{a}^{\dag} (\tau_2)} =
  \frac{1}{Z_S}\int \mathcal{D}[\bar{a} (\tau) a (\tau)] K[\bar{a} (\tau) a(\tau)]
  a(\tau_1) \bar{a}(\tau_2) I[\bar{a}(\tau) a(\tau)].
\end{equation}
It should be noted that if the order of $a (\tau_1)$ and $\bar{a}
(\tau_2)$ changes, then it would generate a minus sign, which corresponds to
the sign definition of in fermion Green's function. To be specific, if
$\tau_2 > \tau_1$ then we need to move $\bar{a} (\tau_2)$ to the left
side of $a (\tau_1)$ for the final Grassmann integral, this yields a minus
sign. There is no sign issue when interchanging $\bar{a} (\tau_2)$ with
other pairs of Grassmann variables.

This expression can be generalized to multiple-time correlation
functions directly. For instance, the four-time correlation function
$\expval*{T \hat{a}(\tau_1) \hat{a} (\tau_2) \hat{a}^{\dag} (\tau_3)
  \hat{a}^{\dag} (\tau_4)}$ can be written as
\begin{equation}
  \expval*{T \hat{a}(\tau_1) \hat{a} (\tau_2) \hat{a}^{\dag} (\tau_3)\hat{a}^{\dag} (\tau_4)} =
  \frac{1}{Z_S} \int \mathcal{D}[\bar{a} (\tau) a (\tau)]
  K[\bar{a} (\tau) a(\tau)] a(\tau_1) a(\tau_2) \bar{a}(\tau_3)\bar{a} (\tau_4) I[\bar{a}(\tau) a(\tau)].
\end{equation}

In Keldysh and Kadanoff formalism, the contour-ordered correlation function
can be expressed in a similar manner. We need to simply insert the operators
at appropriate positions on the contour. For instance, the two-time
contour-ordered correlation function can be written as
\begin{equation}
  \expval*{T_{\mathcal{C}}\hat{a} (t_1) \hat{a}^{\dag} (t_2)} =
  \frac{1}{Z_S} \int\mathcal{D}[\bar{a}(t) a(t)] K[\bar{a}(t) a(t)] a(t_1)\bar{a}(t_2) I[\bar{a}(t) a(t)],
\end{equation}
and the four-time contour-ordered correlation function can be written as
\begin{equation}
  \expval*{T_{\mathcal{C}} \hat{a}(t_1) \hat{a}(t_2) \hat{a}^{\dag} (t_3) \hat{a}^{\dag}(t_4)} =
  \frac{1}{Z_S} \int \mathcal{D} [\bar{a} (t) a (t)] K[\bar{a} (t) a (t)]
  a(t_1) a(t_2) \bar{a}(t_3) \bar{a}(t_4) I[\bar{a}(t) a(t)].
\end{equation}
\section{Generating Functional Method in Fermion Path
Integral}\label{sec:fermion-generating-functional}

\subsection{The coupling energy between the system and environment}

As in the boson case, the system environment coupling energy $\langle
\hat{H}_{S E} \rangle$ in thermal equilibrium can be evaluated via generating
functional method on the imaginary-time axis as
\begin{equation}
  \expval*{\hat{H}_{S E}(\tau_1)} =
  V [\expval*{\hat{a}^{\dag} (\tau_1)\hat{c} (\tau_1)} + \expval*{\hat{c}^{\dag} (\tau_1)\hat{a}(\tau_1)}] =
  2 V \Re \expval*{\hat{a}^{\dag} (\tau_1) \hat{c} (\tau_1)}.
\end{equation}
The quantity $V \expval*{\hat{a}^{\dag} (\tau_1) \hat{c} (\tau_1)}$
contains a system operator and an environment operator. We can insert the
system operator with a source term $\hat{a}^{\dag} e^{- \xi V \hat{c}}$ at
time $\tau = \tau_1$, then the partition function becomes
\begin{equation}
  Z^{\xi} = Z^{(0)}_E \int \mathcal{D} [\bar{a}(\tau) a(\tau)]
  K[\bar{a}(\tau) a(\tau)] \bar{a}(\tau_1) I^{\xi}[\bar{a}(\tau) a(\tau)],
\end{equation}
where
\begin{equation}
  \begin{split}
    I^{\xi} [\bar{a} (\tau) a (\tau)] 
   = & e^{-\int_0^{\beta} \dd{\tau'} \int_0^{\beta}\dd{\tau''}[\bar{a} (\tau') + \xi \delta (\tau'- \tau_1)] \Delta (\tau', \tau'') a (\tau'')}\\
    = & e^{-\int_0^{\beta}\dd{\tau'} \int_0^{\beta} \dd{\tau''}\bar{a}(\tau') \Delta(\tau', \tau'') a(\tau'')
        -\xi\int_0^{\beta}\dd{\tau'} \Delta(\tau_1, \tau') a(\tau')} .
  \end{split}
\end{equation}
The quantity $V \expval*{\hat{a}^{\dag} (\tau_1) \hat{c} (\tau_1)}$
can be evaluated as
\begin{equation}
  V \expval*{\hat{a}^{\dag} (\tau_1) \hat{c} (\tau_1)} =
  - \frac{1}{Z}\eval{\frac{\delta Z^{\xi}}{\delta \xi}}_{\xi = 0},
\end{equation}
which yields
\begin{equation}
  \begin{split}
    V \expval*{\hat{a}^{\dag} (\tau_1) \hat{c} (\tau_1)}
    = & \frac{1}{Z_S} \int \mathcal{D} [\bar{a} (\tau) a (\tau)] K[\bar{a} (\tau) a (\tau)]
        \times \int_0^{\beta} \bar{a} (\tau_1) a(\tau') \Delta (\tau_1, \tau')\\
   = & -\int_0^{\beta} \dd{\tau'} \Delta (\tau_1, \tau') \expval*{T\hat{a} (\tau') \hat{a}^{\dag} (\tau_1)}.
  \end{split}
\end{equation}
Similar to the boson case, on the Keldysh and Kadanoff contour we can
insert a source term $\hat{a}^{\dag}e^{\xi V\hat{c}}$ at $t_1$. Then the
influence functional with a source term is written as
\begin{equation}
  \begin{split}
    I^{\xi}[\bar{a} (t) a (t)]
    = & e^{- \int_{\mathcal{C}} \dd{t'} \int_{\mathcal{C}}\dd{t''}[\bar{a}(t') + i\xi \delta_{\mathcal{C}}(t' - t_1)] \Delta (t', t'') a(t'')}\\
    = & e^{- \int_{\mathcal{C}} \dd{t'} \int_{\mathcal{C}} \dd{t''} \bar{a}(t') \Delta(t', t'') a(t'') - i\xi \int_{\mathcal{C}} \dd{t'}\Delta (t_1, t') a(t')},
  \end{split}
\end{equation}
and
\begin{equation}
  V \expval*{\hat{a}^{\dag} (t_1) \hat{c} (t_1)} =
  \frac{1}{Z} \eval{\frac{\delta Z^{\xi}}{\delta \xi}}_{\xi = 0} =
  i \int_{\mathcal{C}} \dd{t'}\Delta (t_1, t') \expval*{T_{\mathcal{C}}\hat{a} (t') \hat{a}^{\dag} (t_1)}.
  \label{eq:fermion-keldysh-system-environement-coupling}
\end{equation}

\subsection{Currents}

In the fermion case, apart from the energy current, the particle
current is also of great interest in a non-equilibrium setup
\cite{bertrand2019-reconstructing,thoenniss2023-efficient,chen2024-gtempo}. In
the non-equilibrium scenario, the Keldysh formalism is adopted. The
energy current flows out from the environment at time $t_1$ is defined
as
\begin{equation}
  \mathcal{I}(t_1) = - \expval*{\hat{\dot{H}}_E(t_1)} =
  -i\expval*{[\hat{H}, \hat{H}_E (t_1)]} =
  -i V \varepsilon_0 \expval*{\hat{a}^{\dag} (t_1) \hat{c} (t_1) - \hat{c}^{\dag} (t_1) \hat{a} (t_1)} =
  2 V \varepsilon_0 \Im \expval*{\hat{a}^{\dag} (t_1) \hat{c}(t_1)}.
\end{equation}
According to \eqref{eq:fermion-keldysh-system-environement-coupling},
the energy current is immediately evaluated as
\begin{equation}
  \begin{split}
    \mathcal{I} (t_1) 
    = &2\Re \qty[ \int_{\mathcal{C}} \dd{t'} V^2\varepsilon_0 G_0 (\varepsilon_0 ; t_1, t')
        \expval*{T_{\mathcal{C}} \hat{a} (t')\hat{a}^{\dag} (t_1)}]\\
    = &2 \Re \qty[\int_C \dd t' \int \dd{\varepsilon} \Gamma (\varepsilon) \varepsilon
        G_0 (\varepsilon ; t_1, t')\expval*{T_C \hat{a} (t') \hat{a}^{\dag} (t_1)}].
  \end{split}
\end{equation}

If multiple baths are present, then the energy current flows out from
the $\alpha$th bath is
\begin{equation}
  \mathcal{I}_{\alpha} (t_1) = 2 \Re \qty[ \int_{\mathcal{C}} \dd{t'}
  \int\dd{\varepsilon} \Gamma_{\alpha} (\varepsilon) \varepsilon G_{\alpha}(\varepsilon ; t_1, t')
  \expval*{T_{\mathcal{C}} \hat{a} (t') \hat{a}^{\dag} (t_1)}].
\end{equation}

Denote $\hat{N}_E$ as the particle number operator of the
environment. In single fermion case, it is
$\hat{N}_E = \hat{c}^{\dag} \hat{c}$, and in the continuous fermions
case, it is $\hat{N}_E = \sum_k \hat{c}^{\dag}_k \hat{c}_k$. The
particle current is defined as the opposite of change per unit time of
particle number in the environment as
\begin{equation}
  \mathcal{J} (t_1) = - \expval*{\hat{\dot{N}}_E (t_1)} =
  -i\expval*{[\hat{H}, \hat{N}_E(t_1)]} =
  2 V \Im \expval*{\hat{a}^{\dag} (t_1) \hat{c} (t_1)}.
\end{equation}
According to \eqref{eq:fermion-keldysh-system-environement-coupling}, the
particle current expression is
\begin{equation}
  \mathcal{J}(t_1) = 2\Re \qty[ \int_{\mathcal{C}} \dd{t'} \Delta (t_1, t')\expval*{T_{\mathcal{C}}\hat{a} (t') \hat{a}^{\dag} (t_1)}].
\end{equation}
If multiple baths are present, then we have
\begin{equation}
  \mathcal{J}_{\alpha}(t_1) = 2\Re \qty[\int_{\mathcal{C}}\dd{t'}\Delta_{\alpha} (t_1, t')
  \expval*{T_{\mathcal{C}}\hat{a} (t') \hat{a}^{\dag} (t_1)}].
\end{equation}

\subsection{Environment Green's functions}

In imaginary-time formalism, the environment Green's function is defined as
\begin{equation}
  G (\varepsilon_0 ; \tau_1, \tau_2) = - \expval*{T \hat{c}  (\tau_1)\hat{c}^{\dag} (\tau_2)}.
\end{equation}
As in the boson case, we can insert two source terms
$e^{-\xi_1 \hat{c}},e^{-\xi_2 \hat{c}^{\dag}}$ at $\tau_1,\tau_2$, and
the partition function becomes
\begin{equation}
  Z^{\xi_1 \xi_2} = Z^{(0)}_E \int\mathcal{D}[\bar{a}(\tau) a(\tau)]K[\bar{a}(\tau) a(\tau)] I^{\xi_1 \xi_2} [\bar{a}(\tau) a(\tau)],
\end{equation}
where
\begin{equation}
  \begin{split}
    I^{\xi_1 \xi_2} [\bar{a} (\tau) a (\tau)]
    = & e^{- \int_0^{\beta}\dd{\tau'}\int_0^{\beta}\dd{\tau''} [\bar{a} (\tau') +
        \xi_1 V^{-1} \delta (\tau' - \tau_1)] \Delta (\tau', \tau'') [a (\tau'') + \xi_2 V^{-1} \delta (\tau'' - \tau_2)]} \\
    = & e^{-\int_0^{\beta} \dd{\tau'}\int_0^{\beta}\dd{\tau''}\bar{a} (\tau') \Delta (\tau', \tau'') a(\tau'') -
        \xi_1 V^{-1}\int_0^{\beta} \dd{\tau'}\Delta(\tau_1, \tau') a(\tau') -
        \xi_2 V^{-1}\int_0^{\beta} \dd{\tau'}\bar{a}(\tau') \Delta(\tau', \tau_2) -
        \xi_1\xi_2 V^{-2} \Delta(\tau_1, \tau_2)}.
  \end{split}
\end{equation}
Differentiation of $Z^{\xi_1 \xi_2}$ with respect to $\xi_1, \xi_2$ yields
\begin{equation}
  \begin{split}
    G (\varepsilon_0 ; \tau_1, \tau_2)
    = & - \frac{1}{Z} \eval{\frac{\delta^2Z^{\xi_1 \xi_2}}{\delta \xi_1 \delta \xi_2}}_{\xi_1 = \xi_2 = 0}\\
    = & G_0 (\varepsilon_0 ; \tau_1, \tau_2) - V^2\int_0^{\beta}\dd{\tau'}\int_0^{\beta}\dd{\tau''} G_0(\varepsilon ; \tau_1, \tau')
        \expval*{T \hat{a} (\tau') \hat{a}^{\dag} (\tau'')} G_0 (\varepsilon ; \tau'',\tau_2),
  \end{split}
\end{equation}
where $- V^2\expval*{T \hat{a}(\tau') \hat{a}^{\dag}(\tau'')}$ acts as
the $T$-matrix for environment Green's function. In the continuous
fermions case, we have
\begin{equation}
  G (\varepsilon_k ; \tau_1, \tau_2) = G_0 (\varepsilon_k ; \tau_1, \tau_2) -
  V_k^2 \int_0^{\beta} \dd{\tau'} \int_0^{\beta}\dd{\tau''} G_0(\varepsilon_k ; \tau_1, \tau')
  \expval*{T \hat{a} (\tau') \hat{a}^{\dag}(\tau'')}G_0 (\varepsilon_k ; \tau'', \tau_2) .
\end{equation}

In Keldysh and Kadanoff formalism, the environment Green's function is defined
as
\begin{equation}
  G (\varepsilon_0 ; \tau_1, \tau_2) = - i \expval*{T_{\mathcal{C}} \hat{c} (\tau_1) \hat{c}^{\dag} (\tau_2)} .
\end{equation}
Inserting the source terms $e^{\xi_1 \hat{c}},e^{\xi_2 \hat{c}^{\dag}}$
at $t_1,t_2$, we should have
\begin{equation}
  \begin{split}
    I^{\xi_1 \xi_2} [\bar{a}(t) a(t)]
    = & e^{-\int_{\mathcal{C}} \dd{t'}\int_{\mathcal{C}}\dd{t''}
        [\bar{a} (t') + i \xi V^{- 1} \delta_{\mathcal{C}} (t' - t_1)] \Delta(t', t'') [a (t'') +
        i \xi V^{- 1} \delta_{\mathcal{C}} (t'' - t_2)]} \\
    = & e^{- \int_{\mathcal{C}} \dd{t'} \int_{\mathcal{C}} \dd{t''} \bar{a}(t') \Delta(t',t'') a(t'') -
        i\xi_1 V^{-1}\int_{\mathcal{C}} \dd{t'} \Delta (t_1, t') a (t') -
        i\xi_2 V^{-1} \int_{\mathcal{C}} \dd{t'} \bar{a} (t') \Delta(t', t_2)+ \xi_1\xi_2V^{- 2}\Delta(t_1, t_2)},
  \end{split}
\end{equation}
and
\begin{equation}
  \begin{split}
    G(\varepsilon_0 ; t_1, t_2)
    = & - \frac{i}{Z}\eval{\frac{\delta^2Z^{\xi_1 \xi_2}}{\delta \xi_1 \delta\xi_2}}_{\xi_1 = \xi_2 = 0}\\
    = & G_0 (\varepsilon_0 ; t_1, t_2) + i V^{- 2} \int_{\mathcal{C}} \dd{t'} \int_{\mathcal{C}}
        \dd{t''} \Delta (t_1, t') \expval*{T_{\mathcal{C}} \hat{a} (t') \hat{a}^{\dag} (t'')}\Delta (t'', t_2).
  \end{split}
\end{equation}
On the Keldysh contour, the hybridization function $\Delta (t', t'')$ contains
an extra sign ${\mathcal{P}_{t' t''}} $, therefore we have
\begin{equation}
  G (\varepsilon_0 ; t_1, t_2) = G_0 (\varepsilon_0 ; t_1, t_2) -
  i V^2 \int_{\mathcal{C}}\dd{t'} \int_{\mathcal{C}} \dd{t''} \mathcal{P}_{t_1 t'} G_0 (\varepsilon_0 ; t_1,t')
  \expval*{T_{\mathcal{C}} \hat{a} (t') \hat{a}^{\dag} (t'')} \mathcal{P}_{t''t_2} G_0 (\varepsilon_0 ; t'', t_2)
\end{equation}
where
$- i V^2 \mathcal{P}_{t_1 t'} \mathcal{P}_{t'' t_2}
\expval*{T_{\mathcal{C}} \hat{a}(t') \hat{a}^{\dag} (t'')}$ acts as
the $T$-matrix. In continuous fermions case, we have
\begin{equation}
  G (\varepsilon_k ; t_1, t_2) = G_0 (\varepsilon_k ; t_1, t_2) -
  iV_k^2\int_{\mathcal{C}} \dd{t'} \int_{\mathcal{C}} \dd{t''} \mathcal{P}_{t_1 t'} G_0 (\varepsilon_k ;t_1, t')
  \expval*{T_{\mathcal{C}} \hat{a} (t') \hat{a}^{\dag} (t'')}\mathcal{P}_{t'' t_2} G_0 (\varepsilon_k ; t'', t_2).
\end{equation}
On the Kadanoff contour, there is no such an extra sign factor
$\mathcal{P}$, therefore the corresponding expressions is obtained via
removing $\mathcal{P}$ in the above two expressions.

\section{Conclusion}\label{sec:conclusion}

The path integral formalism is a powerful tool for studying quantum
open systems, as it provides a framework to integrate out the
environment degrees of freedom while retaining the effect of the
environment via influence functional. This article gives a
comprehensive derivation of the path integral formalism for both boson
and fermion quantum open systems via coherent states. The influence
functional on the imaginary-time axis, Keldysh and Kadanoff contour
are constructed in a unified framework, where the temporal correlation
in the influence functional can be conveniently expressed by the usual
definition of the Green's function. The generating functional
technique, which allows the retrieval of environment information from
the system correlation functions, is also discussed. Using the
generating functional technique, the expressions for coupling energy
and current between the system and the environment Green's functions
are derived.

For the boson systems, the path integral expression of
Caldeira-Leggett model is derived. For the fermion system, the
Toulouse model is considered and its path integral expression is
derived by introducing the Grassmann algebra.  The derivation shown in
this article can be generalized to other models straightforwardly as
long as the coupling between the system and bath is linear. The path
integral formalism on other contour, such as that in the
out-of-time-ordered correlation function problem
\cite{tuziemski2019-out}, can be derived via the same procedure shown
in this article. The recent development of tensor network methods,
such as TEMPO and GTEMPO, has significantly enhanced the computational
accuracy and efficiency for studying quantum open systems. These
methods enable the direct evaluation of the analytical path integral
expression, and the derivations presented in this article could be
essential for further applications of these methods.

\textbf{Acknowledgment.} This work is supported by the National
Natural Science Foundation of China under Grant No. 12104328.

\appendix

\bibliographystyle{elsarticle-num}
\bibliography{ref}
\end{document}